\documentclass[twocolumn]{aa}
\usepackage{graphicx}
\usepackage{physics,amsmath}
\usepackage{multirow}
\usepackage{txfonts}

\usepackage{natbib}
\bibpunct{(}{)}{;}{a}{}{,} 
\usepackage{xcolor}
\usepackage{caption}
\usepackage{subcaption}
\usepackage{array}
\usepackage[colorlinks=true, allcolors = blue]{hyperref}

\newcolumntype{C}[1]{>{\centering\arraybackslash}p{#1}}

\begin{document} 

   \title{Sub-surface turbulence in hot, massive stars}
   \author{
          C. Van der Sijpt\inst{\ref{inst:KUL}}, F. A.  Driessen\inst{\ref{inst:KUL}},
          J. O. Sundqvist \inst{\ref{inst:KUL}},
          N. Moens \inst{\ref{inst:cmpa}},
          D. Debnath \inst{\ref{inst:KUL}}
          }

   \institute{{Institute of Astronomy, KU Leuven, Celestijnenlaan 200D, 
              3001, Leuven, Belgium \label{inst:KUL}\\
              \email{cassandra.vandersijpt@kuleuven.be}}
        \and {Centre for Mathematical Plasma Astrophysics, Department of Mathematics, KU Leuven, Celestijnenlaan 200B, 3001 Leuven, Belgium 
              \label{inst:cmpa}}}

   \date{Received ; accepted }
   \abstract 
   {Spectroscopic and photometric variability of massive stars indicates the presence of complex, multi-dimensional physics near the stellar surface. Recent advancements in multi-dimensional radiation-hydrodynamical modelling of massive star atmospheres show strong turbulence triggered by the iron opacity bump. }
   {We analyze statistical properties of sub-surface turbulence in multi-dimensional simulations of O-star atmospheres. Particularly, we study potential dependencies of turbulent properties on global stellar parameters.}
    {Using a previously published dataset of two-dimensional O-star models, we studied several statistical properties of the turbulence that develops in these models. We compute autocorrelation functions of density and velocity to study typical eddy length scales. Moreover, we compute kinetic energy power spectra to analyze the energy distribution across spatial scales and compare this to known results from established turbulence theories.}
    {We find anisotropy of the turbulent eddies in velocity space where the radial length dominates. Moreover, this anisotropy increases with classical Eddington factor. This anisotropy disappears in density space where all models show nearly isotropic structures and no dependence on stellar parameters. The kinetic energy power spectra exhibit an inertial range with a cascade exponent that has an average value of $\beta = -2.57$. This exponent does not show any clear dependence on stellar parameters. Importantly, this energy cascade is significantly steeper than those in known turbulence theories such as Kolmogorov turbulence with $\beta = -5/3$ and Burgers turbulence with $\beta = -2$. This additional steepening can potentially be attributed to radiative damping which increases with increasing wavenumber.}   
    {Sub-surface turbulence in hot, massive stars is anisotropic and radially dominated in velocity space and its energy cascade is steeper than both Kolmogorov turbulence and Burgers turbulence. Although not rigorously proven here, we suggest that this steepening is due to the scale-dependent effect of radiative damping.} 

   \title{Properties of near-surface turbulence in hot, massive stars}
   \subtitle{Statistical signatures from multi-dimensional models of O-type stars}
   \authorrunning{C. Van der Sijpt et al.}
   
   \maketitle

    \section{Introduction}
    \label{sec:intro}

    Spectral analysis of hot, massive stars shows indications of a turbulent outer envelope and wind. One major feature is the excessive line broadening that can only be reproduced by the inclusion of so-called `macroturbulence' and `microturbulence' \citep{conti1977,howarth1997,simondiaz2017}. The physics behind this line broadening is, however, poorly understood since traditional massive star atmosphere models are one-dimensional (1D) and computed in radiative equilibrium (see for example: \texttt{FASTWIND} \citealt{santolaya1997,puls2005,sundqvist2018}; \texttt{PoWR} \citealt{grafener2002,hamann2004,sander2012}; \texttt{CMFGEN} \citealt{hillier1998,hillier2001}), and therefore can not self-consistently model multi-dimensional processes such as turbulence. In recent years, there have been major developments in the field of multi-dimensional radiation-hydrodynamical (RHD) atmosphere modelling and these two-dimensional (2D) and three-dimensional (3D) models indeed show significant turbulence near the surface. This turbulence (typically) develops around the iron opacity bump \citep{opal} and is ubiquitous in various types of hot, massive stars: O-stars \citep{jiang2015,schultz2022,debnath2024,delbroek2025,moens2025}, Wolf-Rayet (WR) stars \citep{moens2022} and cooler supergiants \citep{jiang2018,goldberg2026,schillemans2026}. The development of such turbulence is linked to the stellar envelope locally exceeding the Eddington limit, that is, the radiation force can locally overcome gravity. When the Eddington limit is reached, the star will develop a sub-surface convection zone \citep{joss1973,langer1997} and these convection zones have often been linked to macro- and microturbulence \citep{cantiello}. However, in recent work \citep{vandersijpt2025}, we used a linear stability analysis to show that buoyancy-driven convection is not compatible with structure formation in 2D simulations of O-stars and WR stars. Rather, other instabilities, such as Rayleigh-Taylor instabilities and strange mode instabilities, can be triggered by the local increase in opacity near the iron bump \citep{bs03,vandersijpt2025}. Moreover, after the initial structure formation, the turbulent motions are sustained by local differences in opacity and radiative flux (and thus in radiation force) rather than buoyancy \citep{moens2025}. 
    This opacity-driving is very efficient and can lead to supersonic turbulent velocities, on the order of $\sim 100~\mathrm{km/s}$, depending on stellar parameters. Due to these large velocities, the turbulent pressure becomes so substantial that it drastically alters the structure of the stellar envelope \citep{debnath2024,schultz2023}. This turbulent pressure is typically not included in either stellar atmosphere models or stellar structure and evolution models. However, the inclusion of turbulent pressure in stellar atmosphere models has a large impact on the derived stellar parameters \citep{gonzaleztora2025a,gonzaleztora2025b}. While there have been studies on the effect of turbulent pressure in stellar evolution modelling \citep{grassitelli2015}, this has been done with the underlying assumption that the turbulence is driven by sub-sonic convection and they therefore underestimate the resulting velocities. In order to accurately include sub-surface turbulence in 1D stellar structure models, it is important to understand the properties and the underlying physics of the turbulent dynamics. Moreover, while this work focuses on massive stellar envelopes, the results could potentially provide valuable insights for other turbulent systems as well. Turbulence  in radiation-dominated, stratified media has not been extensively studied and massive stellar envelopes provide a useful laboratory to study the underlying physics of turbulence in these environments, which are potentially applicable to other systems with similar physics, such as accretion disks. \\
    \\
    In this paper, we study statistical properties and energy distributions of sub-surface turbulence in O-stars, using 2D models from \citet{moens2025}. Sec. \ref{sec:method} describes the numerical simulation methods and the statistical measures used in this work. In Sec. \ref{sec:results}, we show the results from our analysis and describe how the turbulent properties depend on global stellar parameters. In Sec. \ref{sec:discussion}, we discuss the possible physical mechanism that leads to the results found in Sec. \ref{sec:results} and limitations of our method. Finally, we summarize the main results and describe future research possibilities in Sec. \ref{sec:conclusion}.    
    
    \section{Method}
    \label{sec:method}

    \subsection{Simulation method and model sample}
    \label{sec:simulation}
    The simulations used in this work were performed using the RHD module \citep{moens2021} of the general (radiation-)(magneto)hydrodynamics code \texttt{MPI-AMRVAC} \citep{xia2018,keppens2023}. We solve the multi-dimensional inviscid, compressible Euler equations coupled to a strong radiation field. Therefore, in addition to the Euler equations, we also solve the radiation energy density equation and we account for the radiation force in the momentum equation. These equations are solved on a `box-in-star' Cartesian grid, including a radial dimension (where we correct for spherical divergence; see App. A in \citealt{moens2021}) and one or two lateral dimensions. The gravitational force density $\vec{f}_\mathrm{g}$ of the star is approximated as that of a point-source, so
    \begin{equation}
        \vec{f}_\mathrm{g} = - \rho \frac{G M_\star}{r^2}\vec{\hat{r}},
    \end{equation}
    and the radiation force density is given by 
    \begin{equation}
        \vec{f}_\mathrm{r} = \rho \frac{\kappa_\mathrm{F} \vec{F}}{c}.
    \end{equation}
    Here, $G$ is the gravitational constant, $\rho$ is the gas density, $M_\star$ is the mass of the star, $r$ is the radial coordinate, $c$ is the speed of light, $\vec{F}$ is the radiative co-moving frame flux and $\kappa_\mathrm{F}$ is the flux-weighted mean opacity of the gas. $\kappa_\mathrm{F}$ is calculated following a `hybrid opacity' scheme \citep{poniatowski2022} where $\kappa_\mathrm{F} = \kappa_\mathrm{Ross} + \kappa_\mathrm{line}$, with $\kappa_\mathrm{Ross}$ the Rosseland mean opacity for a static medium and $\kappa_\mathrm{line}$ Doppler-enhanced line opacities in a supersonic medium. In this work, we focus solely on the sub-surface regions where $\kappa_\mathrm{Ross}$, which is calculated from the OPAL opacities \citep{opal}, is the dominant opacity contribution. The radiative flux is calculated from the flux-limited diffusion approximation \citep{pomraning,moens2021}. In the optically thick envelope, this reduces to the well-known diffusion approximation. \\
    \\
    In this work, we study the 2D models presented in \citet{moens2025}. This is a grid of 36 O-stars with classical Eddington parameters in the range $\Gamma_\mathrm{e} \in [0.10,~0.44]$ and masses in the range $M_\star \in [16.6 M_\odot,~68.8 M_\odot]$. They can be split into three populations, with 12 models each: dwarfs, giants and supergiants. All models have a lower boundary which is fixed around a temperature of $T \approx 500~\mathrm{kK}$, which defines the lower boundary radius $R_0$, and they span $2R_0$ radially and $0.4R_0$ laterally. The typical resulting structure of the model is a stably stratified radiative layer near the lower boundary, then a highly turbulent outer envelope around $T \approx 200~\mathrm{kK}$ triggered by the iron opacity bump, and on top of that is a line-driven wind. The grid contains four refinement levels with a cell size of $8 \times 10^{-4}R_0$ at the highest refinement level. \\
    \\
    \citet{moens2025} studied the global turbulent properties of O-stars across this range of stellar parameters. One particular property of interest is the turbulent velocity. It is defined in terms of the turbulent pressure 
    \begin{equation}
        P_\mathrm{turb} = \rho \delta \varv_{r}^2
    \end{equation}
    where $\delta \varv_{r}$ is the radial velocity fluctuation compared to the (laterally averaged) mean velocity, that is, $\delta \varv_{r} = \varv_r - \langle \varv_r \rangle$. 
    The turbulent velocity is then calculated as 
    \begin{equation}
        \varv_\mathrm{turb} = \sqrt{{P_\mathrm{turb}} / \langle \rho \rangle}.
    \end{equation} 
    \citet{moens2025} found sub-surface turbulent velocities in the range $\varv_\mathrm{turb} \in [8~ \mathrm{km/s}, 160~\mathrm{km/s}]$. As the turbulence in these models is driven by the force balance between gravity and the radiation force, which is set by the Eddington parameter $\Gamma = f_\mathrm{r} / f_\mathrm{g}$, the resulting turbulent velocities in \citet{moens2025} show a clear dependence on $\Gamma_\mathrm{e}$, approximately $\varv_\mathrm{turb} \sim \Gamma_\mathrm{e}^2$. \\
    \\
    In this work, we focus only on the optically thick envelope of the star and particularly on the turbulent zone. Therefore, for each model, we locate the radius $r_\mathrm{peak}$ at which the OPAL opacity peaks (that is, the peak of the iron bump) and we take a cut from the total grid centered around this radius, that is, $r \in [r_\mathrm{peak} - 0.2R_0,r_\mathrm{peak} + 0.2R_0]$ and $y \in [-0.2R_0, 0.2R_0]$. However, as an additional requirement on the radial dimension, we impose that the domain needs to be located fully below the average photosphere, such that $r \in [r_\mathrm{peak} - 0.2R_0, \min(r_\mathrm{peak} + 0.2R_0,R_\mathrm{ph})]$. The boundary conditions of the simulation domain and the potential implications on the final results are discussed in Sec. \ref{sec:limitations}. \\
    \\
    It is important to note that in these models, the background medium is stratified and develops a mean velocity after a certain radius. Since we want to study the turbulent dynamics only, and not the global structure of the stellar envelope, we define the turbulent velocity field components as 
    \begin{equation}
        \delta \varv_i = \varv_i - \langle \varv_i \rangle
    \end{equation}
     and $\delta \varv_\mathrm{tot} = \sqrt{\delta \varv_{r}^2 + \delta \varv_{y}^2}$. Analogously, the turbulent density field is 
     \begin{equation}
         \delta \rho = \rho - \langle \rho \rangle.
     \end{equation}
     The angular brackets denote lateral averaging.
    
    \subsection{Correlation lengths}

    A first relevant statistical measure in turbulent flows is the correlation length. To calculate the correlation lengths, we compute autocorrelation functions of certain gas properties, specifically density and velocity. Take any general gas property at a fixed time $f(\vec{x})$, then the normalized two-point autocorrelation function is defined as 
    \begin{equation}
        \centering
        R(\vec{\ell}) = \frac{\sum_{i} f(\vec{x_i}) f(\vec{x}_{i} + \vec{\ell)}}{\sum_i f(\vec{x}_i)^2}.
    \end{equation}
    Note that we implicitly assume that the turbulent flow is statistically homogeneous by summing over all positions $\vec{x}_i$ in the domain. The correlation function is a function of the separation vector $\vec{\ell}$, which can be decomposed into a radial and lateral component, $\vec{\ell} = \ell_r \hat{\vec{r}} + \ell_y \hat{\vec{y}}$.
    The separate radial and lateral length scales can be obtained by taking a 1D slice from the grid and only considering fully radial or lateral separation vectors, such that
    \begin{align}
        && R(\ell_{{r}}) = \Bigg\langle{\frac{\sum_{i} f(r_i, y_0) f(r_i + \ell_{r}, y_0)}{\sum_i f(r_i,y_0)^2} \Bigg\rangle_{y_0, t}}, \\
        && R(\ell_{{y}}) = \Bigg\langle\frac{\sum_{i} f(r_0, y_i) f(r_0, y_i + \ell_y)}{\sum_i f(r_0,y_i)^2} \Bigg\rangle_{r_0, t},
    \end{align}
    where we average over all lateral or radial positions and multiple timesteps. The correlation length is formally defined as 
    \begin{equation}
        \ell_{\mathrm{corr},i} = \int_0^{\infty} R(\ell_i) d \ell_i.
    \end{equation}
    However, assuming the autocorrelation function is a decaying exponential $R(\ell_i) \sim e^{- A \ell_i}$, then the correlation length becomes $\ell_{\mathrm{corr},i} = 1/A$. If $\ell_i = 1/A$, then $R(1/A) = e^{-1}$, so the correlation length can be determined as the point where the autocorrelation function is equal to $1/e$.  

    \subsection{Energy power spectra}

    Another important signature of turbulence is the kinetic energy distribution across length scales. For isotropic turbulence, this is quantified using the turbulent kinetic energy power spectrum which is defined as 
    \begin{equation}
        E(k) = \frac{1}{2} \iint \hat{\vec{\delta \varv}}^{\ast}(\vec{k}) \cdot \hat{\vec{\delta \varv}}(\vec{k}) \ d \mathcal{S}(k),
    \end{equation}
    where $\hat{\vec{\delta \varv}}$ is the Fourier transform of the turbulent velocity field, $\hat{\vec{\delta \varv}}^{\ast}$ is its complex conjugate, $\vec{k}$ is the wavenumber vector and $\mathcal{S}(k)$ denotes the surface of equal $k$ in wavenumber space, centered around $k = 0$. On a discrete 2D grid, this integral then becomes
    \begin{equation}
        E(k) = \pi k \langle \hat{\vec{\delta \varv}}^{\ast}(\vec{k}) \cdot \hat{\vec{\delta \varv}}(\vec{k}) \rangle,
        \label{eq:ekin_spectrum_2D}
    \end{equation}
    where the averaging is done over a circle of radius $k$. In 3D, the power spectrum is
    \begin{equation}
        E(k) = 2 \pi k^2 \langle \hat{\vec{\delta \varv}}^{\ast}(\vec{k}) \cdot \hat{\vec{\delta \varv}}(\vec{k}) \rangle,
        \label{eq:ekin_spectrum_3D}
    \end{equation}
    where the averaging is now done on spheres with radius $k$. Such a velocity power spectrum is characterized by an `inertial range', which is a range in spatial scales where the energy transfer in the turbulent flow is not affected by the forcing nor the viscous dissipation. Instead, the inertial range is characterized by an `energy cascade', where energy is transferred from large eddies to smaller eddies through local interactions. The resulting power spectrum shows power-law behaviour in this inertial range, namely
    \begin{equation}
        E(k) \sim k^{\beta}.
    \end{equation}
    The power-law exponent holds information about the transfer of energy through scales and is an important characteristic of a turbulent flow.
    In classical Kolmogorov turbulence \citep{kolmogorov1941}, which describes incompressible and isotropic turbulence, this exponent is $\beta = -5/3$. In Burgers turbulence \citep{burgers1948}, which describes compressible shock-driven turbulence, the exponent becomes $\beta = -2$. Simulations of supersonic turbulence in the interstellar medium also seem to reproduce $\beta \approx -2$ (for example, \citealt{kritsuk2007}; \citealt{schmidth2009}; \citealt{federrath2021}).
    
    \section{Results}
    \label{sec:results}

    \subsection{Representative model}
    \label{sec:representative}
    We first discuss specific results for a representative model from the total sample: model `G9' from \citet{moens2025}. The relevant stellar parameters of this model are $M = 39.1 M_\odot$, $T_\mathrm{eff} = 34~\mathrm{kK}$, $\varv_\mathrm{turb} \approx 100~\mathrm{km}/\mathrm{s}$ and $\Gamma_\mathrm{e} = 0.35$. This model was chosen because of its significant turbulent velocity, but it still lies approximately in the middle of the range of $\Gamma_\mathrm{e}$.

    \begin{table*}[t]
    \centering
    \caption{Global statistics at various points in time. From left to right, the columns show time (in units of dynamical timescales), the maximum density fluctuation relative to the average density structure, the root-mean-squared Mach number $\mathcal{M} = \sqrt{\langle (\delta \varv_\mathrm{tot}/c_\mathrm{s})^2 \rangle}$, the probability $P(\mathcal{M} > 1)$ and the small-scale compressive ratio $r_\mathrm{cs}$.}
    \label{tab:global_stats}
    \renewcommand{\arraystretch}{1.0}
    \begin{tabular}{C{0.1\textwidth} C{0.1\textwidth} C{0.1\textwidth} C{0.1\textwidth} C{0.1\textwidth}}
    \hline\hline
    \centering Time & \centering $(\delta\rho/ \langle \rho \rangle)_\mathrm{max}$ & \centering $\mathcal{M}$ & \centering $P(\mathcal{M}>1)$ & \centering\arraybackslash $r_\mathrm{cs}$ \\
    \hline\hline
     $30.0$ & $20.74$ & $2.23$ & $0.55$  & $0.48$ \\
     $34.5$ & $17.56$ & $1.67$ & $0.49$ & $0.47$ \\ 
     $39.5$ & $34.11$ & $3.43$ & $0.64$ & $0.57$ \\ 
     $44.5$ & $14.28$ & $2.91$ & $0.54$ & $0.81$ \\ 
     $49.5$ & $19.29$ & $3.05$ & $0.70$ & $0.50$ \\ 
     $54.5$ & $20.07$ & $2.40$ & $0.58$ & $0.43$ \\ 
     $59.5$ & $11.3$ & $1.91$ & $0.54$ & $0.48$ \\ 
     $69.0$ & $23.86$ & $2.83$ & $0.64$ & $0.44$ \\ 
     $79.0$ & $17.4$ & $2.36$ & $0.60$ & $0.38$ \\ 
     $89.0$ & $19.61$ & $2.36$ & $0.62$ & $0.45$ \\ 
     $99.0$ & $14.67$ & $1.63$ & $0.48$ & $0.27$ \\ 
     $109.0$ & $27.16$ & $2.73$ & $0.67$ & $0.53$ \\ 
     $119.0$ & $13.96$ & $2.56$ & $0.60$ & $0.50$ \\ 
     $129.0$ & $15.08$ & $2.54$ & $0.59$ & $0.59$ \\ 
     $139.0$ & $11.87$ & $2.31$ & $0.63$ & $0.38$ \\ 
    \hline\hline
    \end{tabular}
    \end{table*}

    \subsubsection{Global statistics}
    
    Table \ref{tab:global_stats} shows global statistical properties of the turbulent region throughout time. The first property that we studied is the maximum density fluctuation relative to the average density structure at that point. The maximum density fluctuation varies roughly between $15$ and $35$ with stochastic variation through time. The third column shows the average Mach number of the turbulent velocity field, calculated as $\mathcal{M} = \sqrt{\langle (\delta \varv_\mathrm{tot}/c_\mathrm{s})^2\rangle}$ with $c_\mathrm{s} = \sqrt{p / \rho}$ the isothermal gas sound speed. We notice that the Mach number typically fluctuates between $2$ and $3$, meaning that the turbulence is clearly supersonic compared to the gas sound speed. The peaks in Mach number also typically coincide with peaks in density fluctuation. This indicates the creation of shocks which lead to heavy compression of the gas near the shock fronts. We do note, however, that the gas does remain mostly subsonic when comparing to the radiation sound speed $c_\mathrm{r} = \sqrt{(4 E)/(9 \rho)}$, with a Mach number between $\mathcal{M} \sim 0.5-0.9$. We find that typically between $50\%$ and $60\%$ is supersonic compared to the gas sound speed. For the radiation sound speed, however, this is only about $10\%$. The final column shows the small-scale compressive ratio \citep{kida1990} 
    \begin{equation}
        r_\mathrm{cs} = \frac{\abs{\div{\delta \vec{\varv}}}^2}{\abs{\div{\delta \vec{\varv}}}^2 + \abs{\curl{\delta \vec{\varv}}}^2}.
    \end{equation}
    This quantity describes the relative importance of the solenoidal and dilatational modes of the velocity fluctuations. In Table \ref{tab:global_stats}, we see that, on average, $r_\mathrm{cs} \sim 0.48$, with occasional peaks. Although not shown in the table, as turbulence develops during the earlier stages of the simulation, $r_\mathrm{cs}$ starts out as close to zero and gradually increases to finally fluctuate around the average value given above. Therefore, the initial instabilities that trigger structure formation lead to a mostly solenoidal velocity field. However, as the velocity fluctuations grow to supersonic values, the dilatational component becomes more important.
    During steady-state, the dilatational and solenoidal components are comparable. This implies a clear departure from incompressible conditions and shocks can become dynamically important in the turbulent flow, which is consistent with the supersonic Mach numbers found above.
    
    \subsubsection{Autocorrelation}

    \label{sec:rep_autocorrelation}

\begin{figure}
        \centering
        \includegraphics[width=\linewidth]{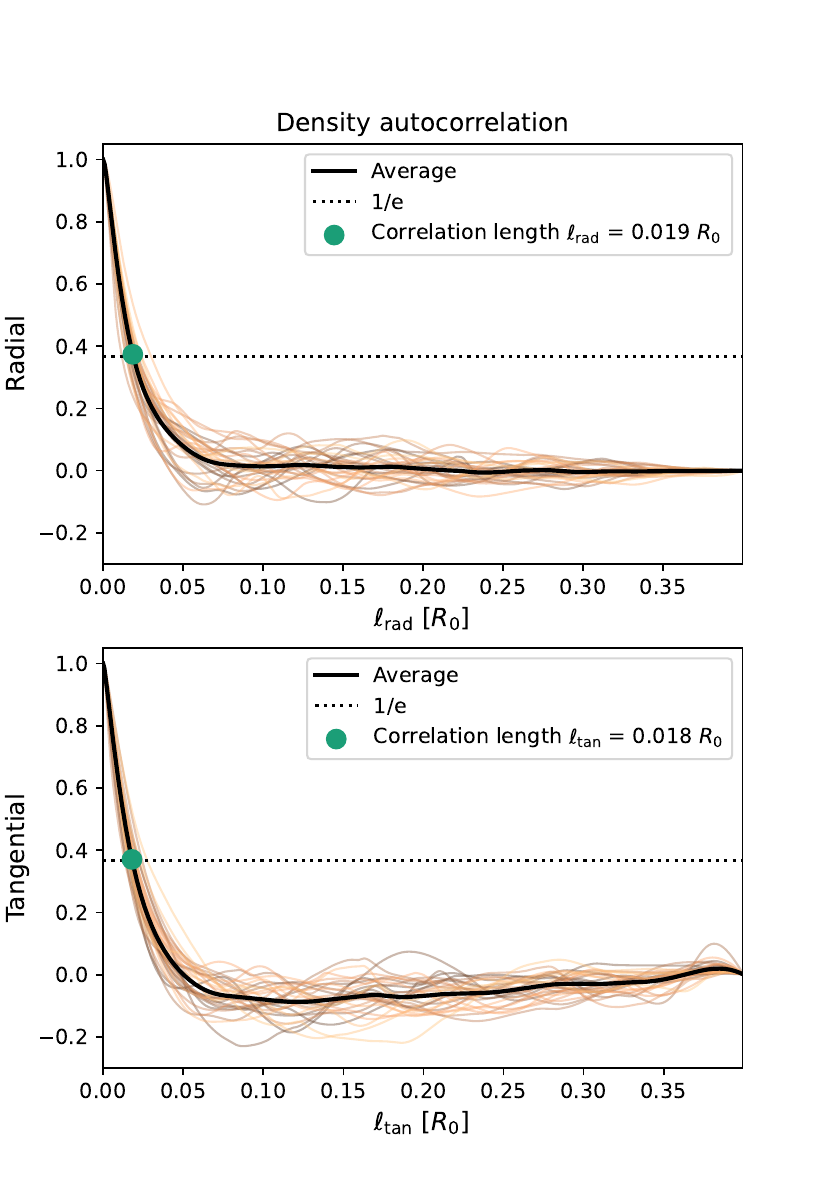}
        \caption{Density autocorrelation functions for model G9. Top panel is the autocorrelation from radial slices and the bottom panel is from tangential slices. The dotted lines indicate the $1/e$ point and the green dots are the derived correlation lengths.}
        \label{fig:rhocorr_star74}
    \end{figure}

\begin{figure}
        \centering
        \includegraphics[width=\linewidth]{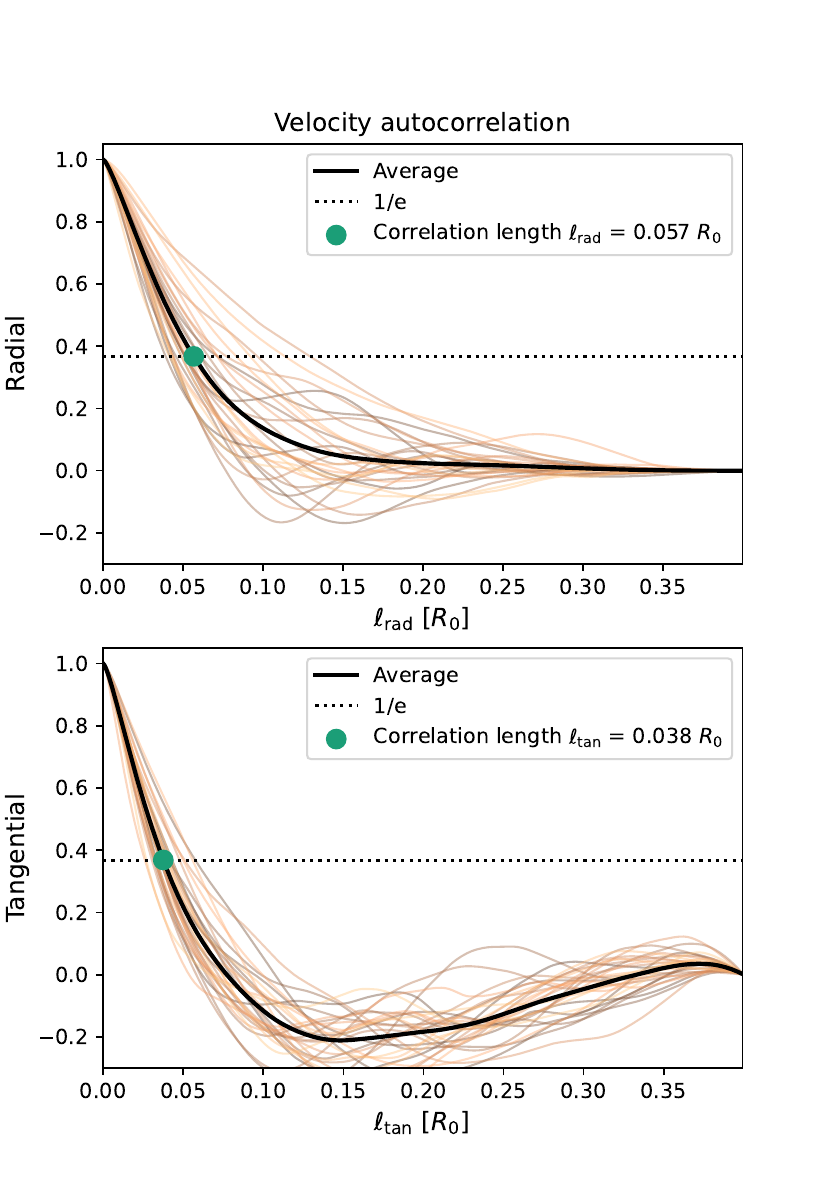}
        \caption{Velocity autocorrelation functions for model G9. Top panel is the autocorrelation of radial velocity, computed over radial slices, and the bottom panel is the autocorrelation of tangential velocity, computed over tangential slices. The dotted lines indicate the $1/e$ point and the green dots are the derived correlation lengths.}
        \label{fig:vel_corr_G9}
    \end{figure}

    In Fig. \ref{fig:rhocorr_star74}, both the radial (top panel) and the lateral (bottom panel) autocorrelation function of $\delta \rho$ are shown for model G9. The black line shows the average autocorrelation function over thirty snapshots and the copper lines show the results for the individual snapshots. The dotted line corresponds to a value of $1/e$ and its intersection with the autocorrelation function determines the correlation length, indicated by the green dot. The radial and the lateral density correlation lengths are $\ell_\mathrm{r} = 0.019~R_0$ and $\ell_\mathrm{y} = 0.018~R_0$, respectively, which shows that a coherent density structure is approximately isotropic. 
    Similarly, Fig. \ref{fig:vel_corr_G9} shows the velocity autocorrelation functions for the same model. We computed the longitudinal correlation functions which means that $\delta \varv_\mathrm{r}$ and $\delta \varv_\mathrm{y}$ were used to compute the radial and the lateral autocorrelation function, respectively. These longitudinal correlation functions are equal for isotropic
    turbulence \citep{pope2000}. The radial and the lateral velocity correlation lengths are $\ell_\mathrm{r} = 0.057~R_0$ and $\ell_\mathrm{y} = 0.038~R_0$, respectively. Therefore, in this model, the turbulent velocity field is dominated by its radial correlation. Furthermore, the structures are significantly smaller in density space than in velocity space. Additionally, the variation in autocorrelation functions through time is also larger for the velocity field. The radial velocity autocorrelation function, in particular, shows broad wings for certain snapshots which indicates that the structures can extend up to half of the total domain size. The correlation length is limited by the simulation domain size for those snapshots, but since only a limited number of snapshots exhibit these broad-winged autocorrelation functions, the averaged correlation length is not significantly impacted by this effect. Finally, we note that the typical correlation lengths are at least a factor of $\sim 20$ larger than the cell size so these turbulent eddies are well resolved in the simulation. 

    \subsubsection{Kinetic energy power spectrum}

    \label{sec:rep_ekin}
    
    Fig. \ref{fig:spectrum_G9} shows the kinetic energy power spectrum, as defined in Eq. \eqref{eq:ekin_spectrum_2D}, for model G9. Similarly to the autocorrelation functions, the average spectrum is shown in black and the temporal variation can be seen from the gray curves. The power spectrum shows a plateau at small wavenumbers which means that most of the energy is contained in the largest scales. At intermediate wavenumbers, the power spectrum shows power-law behaviour in the inertial range. This inertial range is set by the correlation scale of the turbulent velocity field and by the dissipation scale. The dissipation scale is the scale at which kinetic energy starts being transformed into heat through viscosity. Since we model the inviscid Euler equations, the role of viscosity is played by numerical dissipation. In App. \ref{app:visc}, we describe a 1D test case which we used to determine the numerical dissipation scale. We determined the dissipation wavenumber to be $k_\mathrm{diss} \approx 230 / R_0 $. The correlation wavenumber is calculated using the radial velocity correlation length since this is the dominant scale of the problem. To determine the energy cascade in the inertial range, we fit a power-law to the spectrum, which is shown as the dashed pink curve in Fig. \ref{fig:spectrum_G9}. The power-law exponent determined for this model is $\beta = -2.57$. This cascade is significantly steeper than the cascade in Kolmogorov turbulence. Interestingly, the cascade is also steeper than expected from Burgers turbulence.  This result suggests more removal of turbulent kinetic energy towards smaller scales than expected for a purely shock-dominated (Burgers) cascade. This is most likely due to additional physical processes in our model from the inclusion of a strong radiation field, such as the effect of radiative damping (see also Sec. \ref{sec:steepening}) which dissipates kinetic energy more efficiently.

    \subsubsection{Density-weighted power spectrum}
    \label{sec:rep_ekin_dens}
    
\begin{figure}
        \centering
        \includegraphics[width=\linewidth]{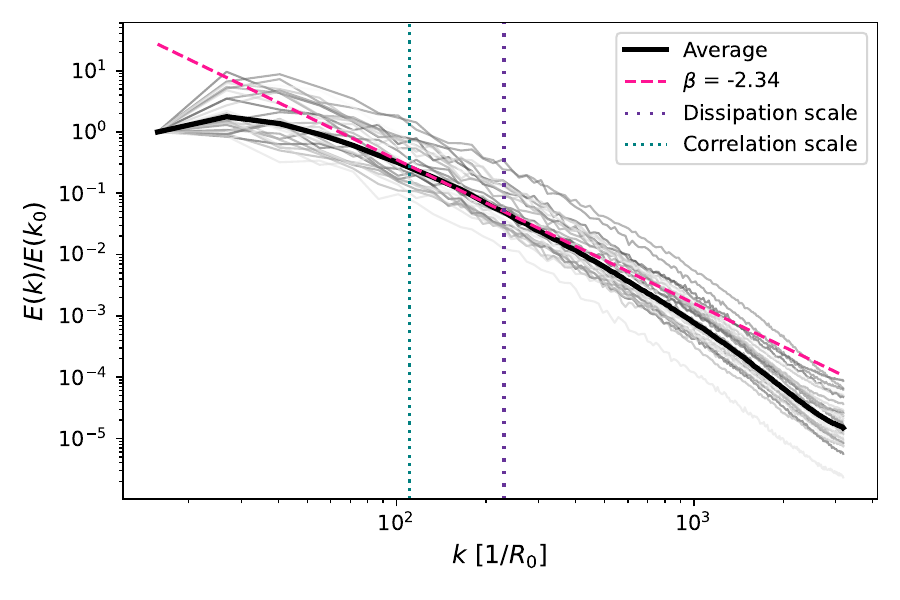}
        \caption{Normalized density-weighted energy power spectrum as a function of wavenumber. The correlation scale and the dissipation scale are indicated by the vertical dotted lines. The dashed line shows a power-law fit to the inertial range of the resulting spectrum.}
        \label{fig:spectrum_G9_dens}
    \end{figure}

    The definition of the kinetic energy power spectrum (Eq. \ref{eq:ekin_spectrum_2D}) comes from the theory of incompressible turbulence. However, it has been suggested that a density-weighted kinetic energy power spectrum might be a more representative statistic for compressible turbulence \citep{kida1990,kritsuk2007}. Then, the power spectrum $\mathcal{E}(k)$ is defined as a function of the variable $w = \sqrt{\rho} \vec{\varv}$ such that
    \begin{equation}
        \mathcal{E}(k) = \pi k \langle \hat{\vec{w}}^{\ast}(\vec{k}) \cdot \hat{\vec{w}}(\vec{k}) \rangle.
        \label{eq:ekin_spectrum_dens}
    \end{equation}
    Fig. \ref{fig:spectrum_G9_dens} shows $\mathcal{E}(k)$ for model G9. The derived energy cascade exponent is $\beta_\rho = -2.34$, which is shallower than the cascade exponent for the classical kinetic energy spectrum. This implies that the kinetic energy budget at smaller scales is dominated by high density regions, which is in line with shock formation. Interestingly, the departure from power-law behaviour after the dissipation scale is also more prominent in the density-weighted spectrum. This may indicate that
the density-weighted velocity ﬁeld is more sensitive to numerical smoothing,
possibly because the density contains comparatively more small-scale structure. 

    \subsection{Analysis of grid of models}

    The methods and results outlined in Sec. \ref{sec:representative} can be extended to the full grid of models in \citet{moens2025}. All time averages are computed over thirty snapshots. As in Sec. \ref{sec:representative}, the turbulent domain was selected for each model by locating the peak of the iron bump at one snapshot and taking a symmetric cut of the full domain around this point (see Sec. \ref{sec:simulation} for details).  

    \subsubsection{Autocorrelation}
\begin{figure}[t!]
        \centering
        \includegraphics[width=\linewidth]{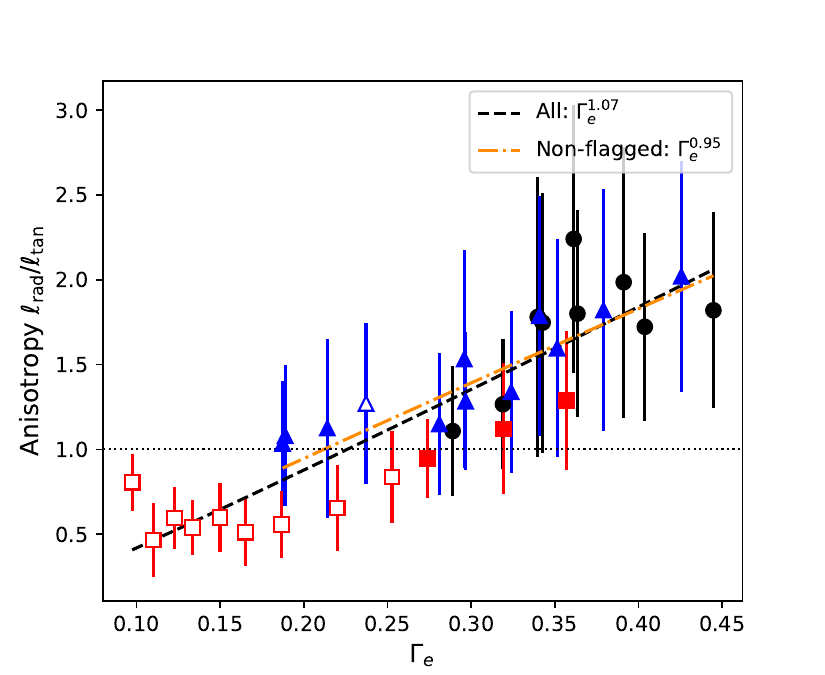}
        \caption{Velocity isotropy factor as a function of $\Gamma_\mathrm{e}$. The colors and markers represent the different populations within the grid, adopted from \citep{moens2025}: red are dwarf stars, blue are giants and black are supergiant. The models for which a limited radial domain was used are flagged by the hollow markers. The dashed black line is a fit through all data points and the orange dot-dashed line is a fit through the non-flagged data points. 
        }
        \label{fig:fiso_vel_gammae}
    \end{figure}

    Density and velocity autocorrelation functions were computed for all models in the grid. Analogously to Sec. \ref{sec:rep_autocorrelation}, the radial and lateral autocorrelation lengths could then be determined. One particular characteristic that is very interesting is the isotropy factor $f = \ell_\mathrm{r} / \ell_\mathrm{y}$ which contains information about the shape of the turbulent structures. This isotropy factor can be calculated both in density space and in velocity space. Fig. \ref{fig:fiso_vel_gammae} shows the calculated velocity isotropy factors as a function of $\Gamma_\mathrm{e}$. Some of the models were cut radially to a domain size smaller than the standard $0.4~R_0$. These models are indicated in the figure with the hollow markers. Although qualitative inspection of the individual autocorrelation functions shows no clear signs of unresolved radial structures (seen by, for example, broad wings in the autocorrelation functions), the radial correlation length might still be impacted by this altered domain. The errorbars are computed from the standard deviation in the correlation lengths over time. A fit to the data, both including all data points and only the non-flagged ones, shows an approximately linear dependence of $f_\mathrm{vel}$ on $\Gamma_\mathrm{e}$. While the lateral correlation lengths stay roughly constant as a function of $\Gamma_\mathrm{e}$, the radial lengths increase with $\Gamma_\mathrm{e}$, leading to the increase in $f_\mathrm{vel}$. We see that most models have turbulent structures which are more radially extended. However, the models with the lowest $\Gamma_\mathrm{e}$ appear to be laterally dominated. We do note that these models have a limited radial domain which might impact these results. Moreover, \citet{moens2025} point out that these models mainly experience global oscillations rather than turbulent motions.  Contrasting the result for the velocity correlation lengths, the turbulent eddies remain nearly isotropic in density space with $f_\mathrm{dens} \approx 1$ for all models.
    

    \subsubsection{Kinetic energy power spectrum}
    \label{sec:ekin_spectra}

\begin{figure}[t!]
        \centering
        \includegraphics[width=\linewidth]{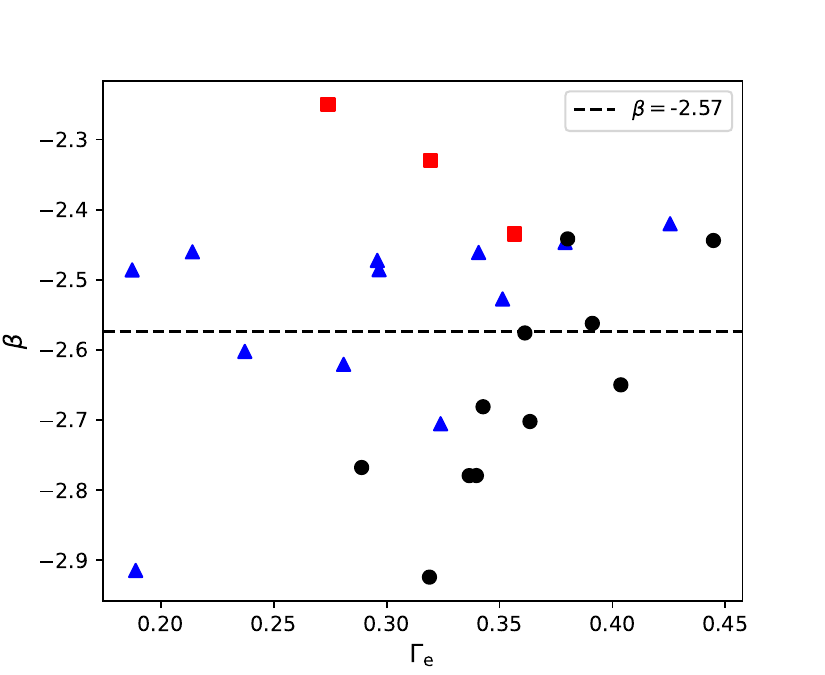}
        \caption{Energy cascade exponents as a function of $\Gamma_\mathrm{e}$. The colors and markers represent the different populations within the grid, adopted from \citep{moens2025}: red are dwarf stars, blue are giants and black are supergiant. The dashed line indicates the average value over the full grid.}
        \label{fig:beta_gammae}
    \end{figure}

    The kinetic energy power spectrum was computed for the full grid of models. Analogously to the procedure described in Sec.~\ref{sec:rep_ekin}, the energy cascade exponents were computed by fitting a power-law to the inertial range for each model. However, in some models, the correlation wavenumber is actually larger than the dissipation wavenumber, meaning that no inertial range could be resolved. Fig. \ref{fig:beta_gammae} shows the resulting cascade exponents as a function of $\Gamma_\mathrm{e}$, for those models with a resolved inertial range. No clear dependence of $\beta$ on $\Gamma_\mathrm{e}$ is found. Rather, the values seem to scatter around the average value of $\beta = -2.57$ and all exponents are steeper than Burgers result of $\beta = -2$.

    \subsubsection{Density-weighted power spectrum}

\begin{figure}[t!]
        \centering
        \includegraphics[width=\linewidth]{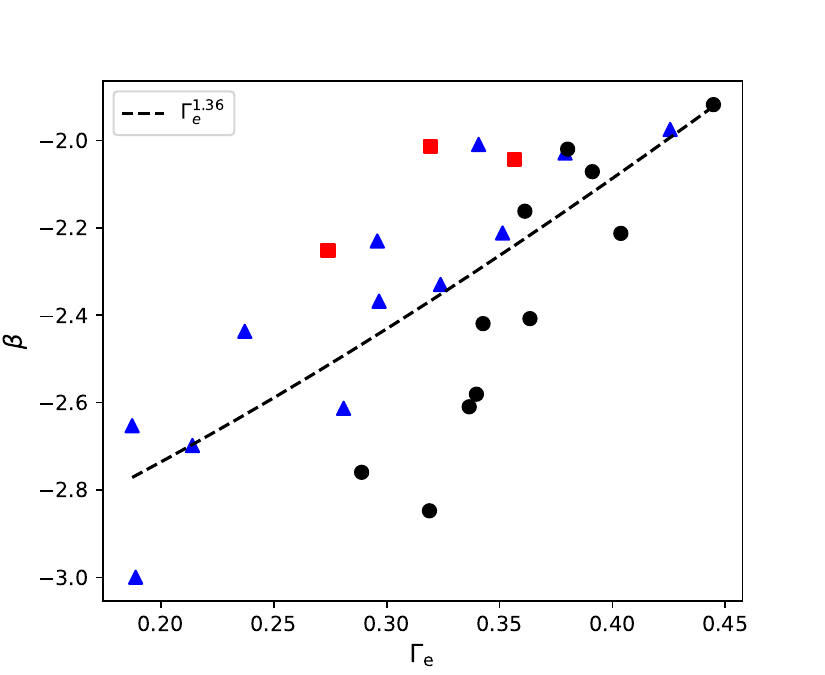}
        \caption{Energy cascade exponents of density-weighted power spectra as a function of $\Gamma_\mathrm{e}$. The colors and markers represent the different populations within the grid, adopted from \citep{moens2025}: red are dwarf stars, blue are giants and black are supergiant. The dashed line shows a fit through the data points.}
        \label{fig:beta_gammae_dens}
    \end{figure}

    The cascade exponents of the density-weighted power spectra, defined in Eq. \eqref{eq:ekin_spectrum_dens}, are shown in Fig. \ref{fig:beta_gammae_dens}. The energy cascade in the density-weighted power spectra does show a dependence on $\Gamma_\mathrm{e}$. The dashed line in the figure shows a fit $\beta \sim \Gamma_\mathrm{e}^{1.36}$ through the data. So, the energy cascade becomes more shallow for larger $\Gamma_\mathrm{e}$. This is related to the degree of compressibility across the grid of models. Models with higher $\Gamma_\mathrm{e}$ show larger density fluctuations on average, that is, the turbulence is more compressible. Therefore, the thin, high-density regions have a larger effect on the density-weighting in the power spectrum at smaller scales, flattening the slope. 
    
    \section{Discussion}
    \label{sec:discussion}

    \subsection{Steepening of energy cascade}
    \label{sec:steepening}
    As discussed in Sec. \ref{sec:ekin_spectra}, the cascade exponents of the kinetic energy power spectrum defined by Eq. \eqref{eq:ekin_spectrum_2D} are steep, with an average of $\beta = -2.57$. Particularly, these exponents are steeper than the typical exponent for Burgers turbulence which is $\beta = -2$. Burgers turbulence is driven by shock formation and is therefore a benchmark model for highly compressible turbulence. Since the turbulence presented here also shows highly compressible filaments, it is noteworthy that the cascade exponents in these models are consistently steeper than expected for purely shock-driven turbulence. This steepening of the energy cascade is likely connected to the strong radiation field that is coupled to the dynamics of the shocks \citep{blaes2011}. As the radiation diffusion time is 
    \begin{equation}
        t_\mathrm{diff} = \frac{3\kappa\rho}{c k^2} = \frac{1}{D k^2},
    \end{equation}
    radiation diffuses more quickly at smaller scales (that is, higher $k$). Therefore, due to rapid radiative diffusion at smaller scales, energy is transported out of a compressed turbulent filament which then becomes damped. Since this damping is stronger at smaller scales, this naturally leads to a steeper energy cascade. 
    \begin{figure}[t!]
        \centering
        \includegraphics[width=\linewidth]{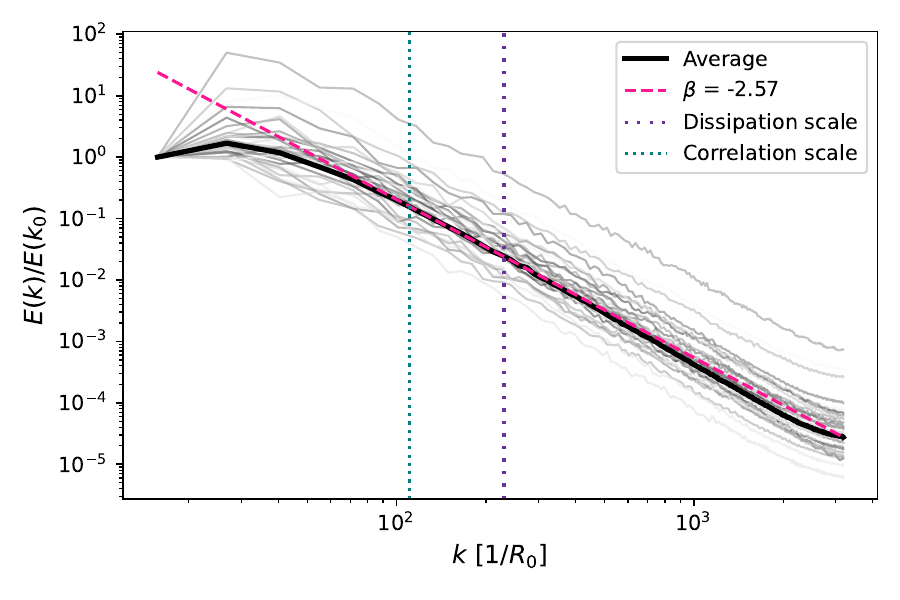}
        \caption{Normalized energy power spectrum as a function of wavenumber. The correlation scale and the dissipation scale are indicated by the vertical dotted lines. The dashed line shows a power-law fit to the inertial range of the resulting spectrum.}
        \label{fig:spectrum_G9}
    \end{figure}

\begin{figure}
        \centering
        \includegraphics[width=\linewidth]{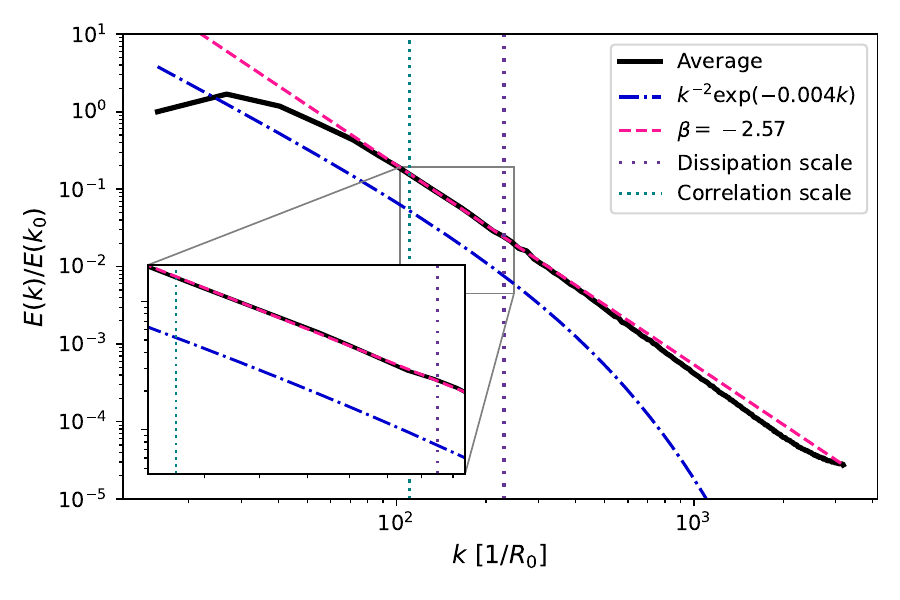}
        \caption{Average energy power spectrum of model G9. The dotted lines represent the correlation and dissipation scales. The dashed pink curve is a power-law fit over the inertial range. The blue dot-dashed curve shows the radiatively damped inertial range power spectrum as derived in the text.}
        \label{fig:power_expcomp}
    \end{figure}

    We can make a more quantitative estimate of how radiative diffusion changes the energy cascade through a scaling analysis. Consider the turbulent velocity field $\delta \vec{\varv}$ which is now affected by radiative damping. This modifies the Fourier amplitude as 
    \begin{equation}
        \hat{\delta \vec{\varv}}(\vec{k},t) = e^{-\omega_\mathrm{rad}t} \hat{\vec{\delta \varv}}_0(\vec{k}),
    \end{equation}
    where $\omega_\mathrm{rad} = D k^2$ is the radiative damping rate and we have defined $\hat{\vec{\delta \varv}}_0(\vec{k})$ to be the Fourier velocity amplitude without the effect of radiative diffusion. We now assume that, when excluding radiative diffusion, the turbulence would be determined entirely by shock generation. This assumption defines the relevant timescale that needs to be compared to the radiative diffusion timescale, namely the time it takes for a shock to compress the gas. This compression time is $t_\mathrm{comp} = 2 \pi (\varv_\mathrm{s} k)^{-1}$, with $\varv_\mathrm{s}$ the typical shock propagation velocity. Our assumption also implies that $\abs{\hat{\vec{\delta \varv}}_0(k)}^2 \sim k^{-2}$, the expected scaling for Burgers turbulence. Thus, the total Fourier amplitude now becomes
    \begin{equation}
        \hat{\delta \vec{\varv}}(\vec{k},t) = e^{-\omega_\mathrm{rad}t_\mathrm{comp}} \hat{\vec{\delta \varv}}_0(\vec{k}) = e^{- 2 \pi D k/\varv_\mathrm{s}} \hat{\vec{\delta \varv}}_0(\vec{k}).
    \end{equation}
    Then, the resulting power spectrum is
    \begin{equation}
        E(k) = e^{-4 \pi D k/\varv_\mathrm{s}} \abs{\hat{\vec{\delta \varv}}_0(k)}^2 \sim e^{-4 \pi D k/\varv_\mathrm{s}} k^{-2}.
    \end{equation}
    Finally, the effective cascade exponent is
    \begin{equation}
        \beta_\mathrm{eff} = \frac{d \ln E(k)}{d \ln k} = -2 - 4 \pi k \frac{D}{\varv_\mathrm{s}}.
    \end{equation}
    Therefore, true power-law behaviour is broken as the steepening of the cascade is scale-dependent. However, in the limited inertial ranges from our simulations, this may still appear as a power-law, albeit steeper than for purely shock-driven turbulence. We can estimate the effective cascade exponent for our representative model described in Sec. \ref{sec:representative}. In Fig. \ref{fig:power_expcomp}, we show the average power spectrum of model G9, but now we have also plotted $\exp(-0.004k)k^{-2}$ as the blue line. The factor in the exponent was chosen to qualitatively match the inertial range of the average spectrum. We see that this functional form can indeed appear as an approximate power law in the inertial range, although there is a much steeper drop-off at higher $k$. Therefore, we should find that $ 4 \pi D/\varv_{s} \approx 0.004 R_0$ from the simulations. We can estimate this ratio by using some characteristic values that we have obtained. In Table \ref{tab:global_stats}, we see that the velocity field has a typical Mach number in the range $2-3$. Therefore, we can assume that a typical shock will have a velocity of $\varv_\mathrm{s} = 2.5 c_\mathrm{s} \approx 125~\mathrm{km/s}$. Then, to estimate the diffusion coefficient, we need a typical shock density and opacity. The opacity is on the order of $\kappa = 3\kappa_\mathrm{e} \approx 1~\mathrm{cm}^2/\mathrm{g}$ and in Table \ref{tab:global_stats}, we see that the highest density fluctuations are in the range of $10-20$. Taking a conservative estimate of $\delta \rho/\langle \rho \rangle = 10$ and using a characteristic background density of $\langle \rho \rangle = 10^{-8}~\mathrm{g}/\mathrm{cm}^3$, we find $\delta \rho \approx 10^{-7}~\mathrm{g}/\mathrm{cm}^3$. Then, the diffusion coefficient becomes $D = c/(3 \kappa \rho) \approx 10^{17}~\mathrm{cm}^2/\mathrm{s}$. Finally, we get $4 \pi D / \varv_\mathrm{s} \approx 12 \times 10^{17} / (125 \times 10^{5})~\mathrm{cm}$, or in units of $R_0 \approx 12R_\odot$, $\approx 0.1R_0$. This estimate is two orders of magnitude larger than the qualitative result found above. Of course, this is only a rough estimate and should be improved upon. Even so, the discrepancy is quite large. While we believe the qualitative picture of radiative damping introducing a scale-dependence that becomes stronger for smaller scales will hold, the analysis presented above is too simplistic to give a quantitative description of the turbulent energy cascade. Two of our assumptions, in particular, might be too crude. Firstly, we assumed that the energy spectrum could be described by Burgers result in the limit of $t_\mathrm{diff} \rightarrow \infty$. Secondly, we assumed that radiative damping directly damps the velocity fluctuations. However, in reality, radiative damping will influence the radiation energy density which, in turn, impacts the kinetic energy and thereby the velocity field. A more rigorous analysis should include the coupling between radiation energy and kinetic energy.

    \subsection{Limitations of simulation set-up}
    \label{sec:limitations}
    The set-up of the sub-domain of the total simulation grid described in Sec. \ref{sec:simulation} introduces some uncertainties in the statistical signatures described here. One particular issue is that of boundary conditions. While the lateral dimension has periodic boundary conditions, this is not the case for the radial dimension. This can be an issue for the resulting power spectra as discontinuities at the boundaries can show up as artificial features in the Fourier transform. However, qualitative inspection of the turbulent velocity field at the radial boundaries shows that the discontinuities are only moderate. Furthermore, these discontinuities would impact only the largest wavenumbers which lie outside the inertial range studied here. Therefore, we do not believe that the calculated energy cascade exponents are affected by the radial boundary effects. Another limitation of our sub-domain is that total turbulent kinetic energy is not conserved. Indeed, energy can be advected out of the sub-domain as it gets dragged along with the wind. When calculating the total turbulent kinetic energy, it fluctuates stochastically over time, but it seems that the combination of material being advected out and material falling back keeps the total energy fluctuations reasonably small. Additionally, as the numerical viscosity in these simulations is significant (see App. \ref{app:visc}), future turbulence studies would benefit from running simulations in higher resolution, ideally increasing the resolution by at least a factor four. 
    The inertial range defined in Sec. \ref{sec:rep_ekin} is very narrow and this limits the cascade fitting procedure to only five to ten data points. While qualitative inspection of goodness of fit can still be done, the statistical uncertainty can only be reduced by extending the inertial range. However, \citet{moens2025} and other previous studies \citep{moens2022,debnath2024,delbroek2025,schillemans2026} focussed on global stellar properties for which this resolution was sufficient. Therefore, for this first dedicated turbulence study, we opted to use the previously available data with only a dedicated resolution test (see App. \ref{app:resolution}) which shows that doubling the resolution does not impact the derived inertial range scaling. Finally, the lateral correlation lengths are likely affected by the pseudo-planar geometry that we use as we do not capture spherical expansion effects in the lateral direction. While we believe this effect will not change the overall result of anisotropy, as we consider a relatively small domain, it is still an important caveat to keep in mind.

    \subsection{Effect of anisotropic driving force}
    An important feature of the turbulence in hot, massive stellar envelopes is that the driving force of the turbulence is not at all isotropic. Indeed, the two bulk forces in the system that impact the turbulent dynamics are gravity and the radiation force, both of which act predominantly in the radial direction. 
    Moreover, the strength of both of these forces also varies radially. While the radial and lateral dimension are split up in the analysis of the autocorrelation lengths, this is not the case for the power spectra analysis. Indeed, by averaging the Fourier coefficients over total wavenumber, we lose information about the individual components $k_\mathrm{r}$ and $k_\mathrm{y}$. To study the individual contributions of both dimensions to the final power spectrum, we computed 1D power spectra $E(k_\mathrm{y})$ and $E(k_\mathrm{r})$. Now the Fourier transform is only computed along the lateral or the radial direction, respectively. Computing the 1D spectra independently for each radial and lateral slice also allows us to probe spatial variations in the energy spectrum, particularly in response to variations in the driving force. For the representative G9 model, the average lateral power spectrum $E(k_\mathrm{y})$, averaged over all radii and over time, shows an energy cascade exponent of $\beta = -2.65$. From the radial power spectrum $E(k_\mathrm{r})$, we find a cascade exponent $\beta = -2.44$. Studying $E(k_\mathrm{y})$ as a function of $r$, we see that close to the lower boundary, the power spectrum shows a clear plateau at large scales and a very steep drop-off to smaller scales, meaning that not much energy is contained in small scales. This is a result of the particular sub-domain of the full grid that we study. As mentioned, we consider a symmetric domain around the peak of the iron bump, but this might still include part of the mostly stable radiative layer towards the lower edge of the domain. In this region, the turbulent velocity field is close to zero and any fluctuations are small in amplitude but extend over relatively large scales. Therefore, the resulting power spectrum shows an extended plateau and steep drop-off. The cascade smoothly flattens towards larger radii as the turbulence fully develops. Even at the top of the domain, however, the cascade is still significantly steeper than Burgers turbulence. So, although the average cascade slope is slightly larger due to these lower edge effects, this does not alter the general conclusions about the energy spectrum of the considered turbulence. Moreover, while there are minor differences between the radial and the lateral power spectrum, the overall shape and slope of the two spectra are very similar. Therefore, although the driving force and the correlation lengths show clear anisotropy, it seems that the 1D power spectrum still provides a reasonable characterization of this sub-surface turbulence.

    \subsection{Comparison between 2D and 3D turbulence}

\begin{figure}[h!]
        \centering
        \includegraphics[width=0.9\linewidth]{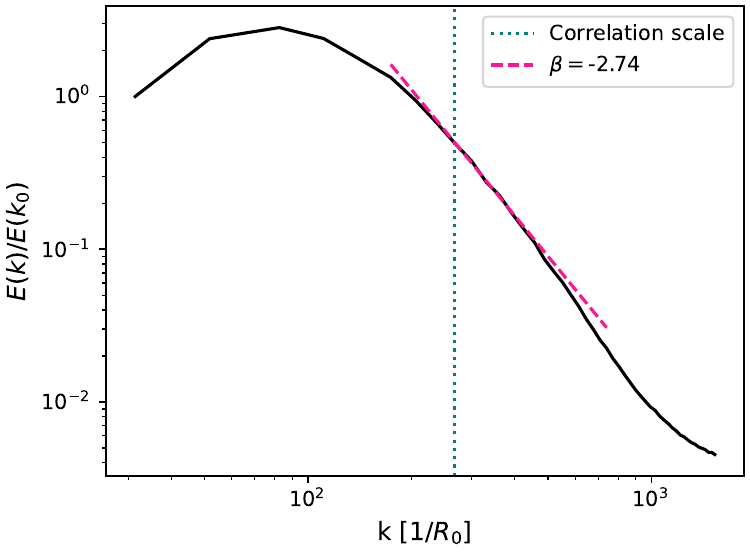}
        \caption{Normalized energy power spectrum of the 3D O4 model from \citet{delbroek2025} as a function of wavenumber. The correlation scale is indicated by the vertical dotted line. The dashed line shows a power-law fit to an approximate inertial range (see text) and the exponent is shown in the legend.}
        \label{fig:power_3D}
    \end{figure}

    The models discussed in Sec. \ref{sec:results} are all 2D models. However, it is well known that the physics of turbulence in 2D is inherently different from turbulence in 3D. Indeed, in 2D, a dual cascade may appear: a forward cascade of enstrophy (which is $\sim \abs{\curl{\vec{\varv}}}^2$) and an inverse cascade of kinetic energy \citep{kraichnan1967,kraichnan1980}. Therefore, the energy in 2D turbulence typically piles up at the largest scales and the flow self-organizes. It must be noted, though, that the leading theory behind this dual cascade is developed for incompressible turbulence. Looking at Fig. \ref{fig:spectrum_G9}, we can not distinguish the two distinct inertial ranges characteristic of the dual cascade. The only deviation from the power-law behaviour in the indicated inertial range can be seen at high wavenumbers, already in the dissipation regime. Moreover, the self-organization of the flow that is typical for an inverse energy cascade should manifest itself as a clear time-dependent evolution of the power spectra. Indeed, the overall power and the plateau at small $k$ would then grow over time. However, qualitatively looking at the spectra from individual timesteps (the gray lines in Fig. \ref{fig:spectrum_G9}), the variation in the spectra is stochastic and shows no clear time-dependence. This seems to indicate a single forward energy cascade. The lack of an inverse cascade is likely linked to the supersonic velocity field in these simulations. Recent work \citep{fouxon2023} studied the possibility of a cascade reversal in 2D turbulence with increasing Mach number and they show that 2D turbulence is very similar to 3D turbulence in the supersonic limit. \\
    Regardless of this, it is still prudent to study how the results in Sec. \ref{sec:results} from 2D models would translate to a physical 3D star. We studied the 3D O-star model described in \citet{delbroek2025}. This is the 3D equivalent of the 2D O4 model in \citet{debnath2024} and we can therefore make a clean one-to-one comparison between this 2D and 3D model. We calculated the velocity correlation lengths for the 3D model and found $\ell_\mathrm{r} \approx 0.023~R_0$, $\ell_\mathrm{y} \approx 0.024~R_0$ and $\ell_\mathrm{z} \approx 0.021~R_0$. For the 2D model, we find $\ell_\mathrm{r} \approx 0.024~R_0$ and $\ell_\mathrm{y} \approx 0.023~R_0$. So, the overall scale of the structures in velocity space agrees well between the 2D and the 3D model. When considering the power spectra (calculated using Eq. \ref{eq:ekin_spectrum_2D} or \ref{eq:ekin_spectrum_3D}, depending on the dimensionality), we found that the resolution of both models is too poor to properly resolve the inertial range. Therefore, we could not get a robust estimate of the power-law exponent in the inertial range. However, we defined an approximate inertial range by starting from the correlation scale, that is taken to be the radial velocity correlation length as before, and estimating the width of the inertial range based on the results in Sec. \ref{sec:results}. The resulting 3D power spectrum and the power-law fit are shown in Fig. \ref{fig:power_3D}. While the overall shape of the spectrum looks similar to that of the 2D model, the 3D spectrum does show a more clearly defined plateau at low $k$. The approximate inertial range has a cascade exponent of $\beta = -2.74$. As a comparison, the equivalent 2D model results in an exponent of $\beta = -2.41$. These exponents agree reasonably well and the finding of steep energy decays emerges in both 2D and 3D.

    \section{Conclusion and future work}
    \label{sec:conclusion}

    In this paper, we studied the statistical properties of sub-surface turbulence in the outer envelopes of O-stars. To do this, we used a grid of 2D radiation-hydrodynamical O-star models. Firstly, we found that, in steady-state, the turbulence is highly supersonic compared to the gas sound speed and the dilatational and solenoidal components are comparable. This means that compressibility, due to shock formation, plays a non-negligible role in turbulent dynamics in these stellar envelopes. Moreover, we quantified typical eddy length scales using autocorrelation functions of velocity and density. We found a clear anisotropy of the turbulent eddies in velocity space, with the radial correlation length dominating. This anisotropy increases linearly with $\Gamma_\mathrm{e}$. In density space, the eddies are nearly isotropic and independent of stellar parameters. Additionally, we computed the turbulent kinetic energy power spectra and we found that the energy cascade exponent $\beta$ is independent of $\Gamma_\mathrm{e}$, with an average of $\beta = -2.57$. This is significantly steeper than the exponent in Kolmogorov turbulence, which is $\beta = -5/3$. Interestingly, Burgers turbulence, the leading theory for shock-driven turbulence, also finds a shallower slope with $\beta = -2$. This means that most of the energy is contained in the largest scales and energy decays to small scales more rapidly than in these established theories. However, if we include a density-weighting in the velocity field, the cascade exponent does depend on $\Gamma_\mathrm{e}$, becoming less steep as $\Gamma_\mathrm{e}$ increases. We hypothesized that the steepening of the energy cascade is linked to radiative damping that acts more strongly on smaller scales. We found scale-dependent radiative damping would actually break the expected power-law behaviour in the inertial range. Instead, we expect a scale-dependent exponent, but a broader inertial range is needed to test this. \\
    In future work, we will conduct a more rigorous analysis of the effect of radiative damping on the turbulent energy cascade. The analysis presented in this work should also be extended to different massive star regimes such as cooler supergiants and lower metallicity environments. Moreover, this work would benefit from a simulation grid at sufficiently high resolution to test the hypothesis of the broken power-law behaviour presented in Sec. \ref{sec:steepening}. Lastly, the results presented in this paper can be used to extend the turbulent pressure formalism implemented in 1D atmosphere models \citep{gonzaleztora2025a,gonzaleztora2025b}.
\begin{acknowledgements}

The computational resources used for this work were provided by Vlaams Supercomputer Centrum (VSC) funded by the Research Foundation-Flanders (FWO) and the Flemish Government. CVdS, JOS and DD acknowledge the support of the European Research Council (ERC) Horizon Europe under grant agreement number 101044048. JOS further acknowledges support from the Belgian Research Foundation Flanders (FWO) Odysseus program under grant number G0H9218N, from FWO grant G077822N, and from KU Leuven C1 grant BRAVE C16/23/009. FAD acknowledges support from FWO grant G0A2O25N. NM acknowledges support from the SPACE CoE, funded by the EU and several partner countries under grant number 101093441. The authors would like to thank all members of the KUL-EQUATION group for fruitful discussion, comments, and suggestions. The following packages were used to analyze the data: WindSimTools \citep{moens2025}, {\fontfamily{qcr}\selectfont NumPy} \citep{harris_2020}, {\fontfamily{qcr}\selectfont SciPy} \citep{virtanen_2020}, {\fontfamily{qcr}\selectfont matplotlib} \citep{hunter_2007}, {\fontfamily{qcr}\selectfont Python amrvac\_reader} \citep{keppens_2020}. 

\end{acknowledgements}

\bibliographystyle{aa}
\bibliography{bibliography.bib} 

@BOOK{mihalas,
       author = {{Mihalas}, D. and {Mihalas}, B.~W.},
        title = "{Foundations of radiation hydrodynamics}",
         year = 1984,
       adsurl = {https://ui.adsabs.harvard.edu/abs/1984frh..book.....M}
}

@ARTICLE{bs03,
       author = {{Blaes}, Omer and {Socrates}, Aristotle},
        title = "{Local Radiative Hydrodynamic and Magnetohydrodynamic Instabilities in Optically Thick Media}",
      journal = {\apj},
         year = 2003,
        month = oct,
       volume = {596},
       number = {1},
        pages = {509-537},
          doi = {10.1086/377637},
archivePrefix = {arXiv},
       eprint = {astro-ph/0304348},
 primaryClass = {astro-ph},
       adsurl = {https://ui.adsabs.harvard.edu/abs/2003ApJ...596..509B}
}

@ARTICLE{jiang2015,
       author = {{Jiang}, Yan-Fei and {Cantiello}, Matteo and {Bildsten}, Lars and {Quataert}, Eliot and {Blaes}, Omer},
        title = "{Local Radiation Hydrodynamic Simulations of Massive Star Envelopes at the Iron Opacity Peak}",
      journal = {\apj},
         year = 2015,
        month = nov,
       volume = {813},
       number = {1},
          eid = {74},
        pages = {74},
          doi = {10.1088/0004-637X/813/1/74},
archivePrefix = {arXiv},
       eprint = {1509.05417},
 primaryClass = {astro-ph.SR},
       adsurl = {https://ui.adsabs.harvard.edu/abs/2015ApJ...813...74J}
}

@ARTICLE{moens2022,
       author = {{Moens}, N. and {Poniatowski}, L.~G. and {Hennicker}, L. and {Sundqvist}, J.~O. and {El Mellah}, I. and {Kee}, N.~D.},
        title = "{First 3D radiation-hydrodynamic simulations of Wolf-Rayet winds}",
      journal = {\aap},
         year = "2022b",
        month = sep,
       volume = {665},
          eid = {A42},
        pages = {A42},
          doi = {10.1051/0004-6361/202243451},
archivePrefix = {arXiv},
       eprint = {2203.01108},
 primaryClass = {astro-ph.SR},
       adsurl = {https://ui.adsabs.harvard.edu/abs/2022A&A...665A..42M}
}

@ARTICLE{debnath2024,
       author = {{Debnath}, D. and {Sundqvist}, J.~O. and {Moens}, N. and {Van der Sijpt}, C. and {Verhamme}, O. and {Poniatowski}, L.~G.},
        title = "{2D unified atmosphere and wind simulations of O-type stars}",
      journal = {\aap},
         year = 2024,
        month = apr,
       volume = {684},
          eid = {A177},
        pages = {A177},
          doi = {10.1051/0004-6361/202348206},
archivePrefix = {arXiv},
       eprint = {2401.08391},
 primaryClass = {astro-ph.SR},
       adsurl = {https://ui.adsabs.harvard.edu/abs/2024A&A...684A.177D}
}

@article{moens2021,
	author = {{Moens}, N. and {Sundqvist}, J. O. and {El Mellah}, I. and {Poniatowski}, L. and {Teunissen}, J. and {Keppens}, R.},
	title = {Radiation-hydrodynamics with MPI-AMRVAC
 - Flux-limited diffusion},
	DOI= "10.1051/0004-6361/202141023",
	url= "https://doi.org/10.1051/0004 -6361/202141023",
	journal = {A\&A},
	year = "2022a",
	volume = 657,
	pages = "A81",
}

@ARTICLE{xia2018,
       author = {{Xia}, C. and {Teunissen}, J. and {El Mellah}, I. and {Chan{\'e}}, E. and {Keppens}, R.},
        title = "{MPI-AMRVAC 2.0 for Solar and Astrophysical Applications}",
      journal = {\apjs},
         year = 2018,
        month = feb,
       volume = {234},
       number = {2},
          eid = {30},
        pages = {30},
          doi = {10.3847/1538-4365/aaa6c8},
archivePrefix = {arXiv},
       eprint = {1710.06140},
 primaryClass = {astro-ph.SR},
       adsurl = {https://ui.adsabs.harvard.edu/abs/2018ApJS..234...30X}
}

@article{poniatowski2022,
	author = {{Poniatowski}, L. G. and {Kee}, N. D. and {Sundqvist}, J. O. and {Driessen}, F. A. and {Moens}, N. and {Owocki}, S. P. and {Gayley}, K. G. and {Decin}, L. and {de Koter}, A. and {Sana}, H.},
	title = {Method and new tabulations for flux-weighted line opacity and radiation line force in supersonic media★},
	DOI= "10.1051/0004-6361/202142888",
	url= "https://doi.org/10.1051/0004-6361/202142888",
	journal = {A\&A},
	year = 2022,
	volume = 667,
	pages = "A113",
}

@ARTICLE{opal,
       author = {{Iglesias}, C. A. and {Rogers}, F. J.},
        title = "{Updated Opal Opacities}",
      journal = {\apj},
         year = 1996,
        month = jun,
       volume = {464},
        pages = {943},
          doi = {10.1086/177381},
       adsurl = {https://ui.adsabs.harvard.edu/abs/1996ApJ...464..943I}
}

@article{cantiello,
	author = {{Cantiello}, M. and {Langer}, N. and {Brott}, I. and {de Koter}, A. and {Shore}, S. N. and {Vink}, J. S. and {Voegler}, A. and {Lennon}, D. J. and {Yoon}, S.-C.},
	title = {Sub-surface convection zones in hot massive stars   and
 their observable consequences*},
	DOI= "10.1051/0004-6361/200911643",
	url= "https://doi.org/10.1051/0004-6361/200911643",
	journal = {A\&A},
	year = 2009,
	volume = 499,
	number = 1,
	pages = "279-290",
}

@ARTICLE{pomraning,
       author = {{Levermore}, C.~D. and {Pomraning}, G.~C.},
        title = "{A flux-limited diffusion theory}",
      journal = {\apj},
         year = 1981,
        month = aug,
       volume = {248},
        pages = {321-334},
          doi = {10.1086/159157},
       adsurl = {https://ui.adsabs.harvard.edu/abs/1981ApJ...248..321L}
}

@ARTICLE{keppens2023,
       author = {{Keppens}, R. and {Popescu Braileanu}, B. and {Zhou}, Y. and {Ruan}, W. and {Xia}, C. and {Guo}, Y. and {Claes}, N. and {Bacchini}, F.},
        title = "{MPI-AMRVAC 3.0: Updates to an open-source simulation framework}",
      journal = {\aap},
         year = 2023,
        month = may,
       volume = {673},
          eid = {A66},
        pages = {A66},
          doi = {10.1051/0004-6361/202245359},
archivePrefix = {arXiv},
       eprint = {2303.03026},
 primaryClass = {astro-ph.IM},
       adsurl = {https://ui.adsabs.harvard.edu/abs/2023A&A...673A..66K}
}

@ARTICLE{conti1977,
       author = {{Conti}, P.~S. and {Ebbets}, D.},
        title = "{Spectroscopic studies of O-type stars. VII. Rotational velocities V sin i and evidence for macroturbulent motions.}",
      journal = {\apj},
         year = 1977,
        month = apr,
       volume = {213},
        pages = {438-447},
          doi = {10.1086/155173},
       adsurl = {https://ui.adsabs.harvard.edu/abs/1977ApJ...213..438C}
}

@article{simondiaz2017,
  title={The IACOB project-III. New observational clues to understand macroturbulent broadening in massive O-and B-type stars},
  author={Sim{\'o}n-D{\'\i}az, Sergio and Godart, M{\'e}lanie and Castro, Norberto and Herrero, Artemio and Aerts, Conny and Puls, Joachim and Telting, J and Grassitelli, Luca},
  journal={A\&A},
  volume={597},
  pages={A22},
  year={2017},
  publisher={Edp Sciences}
}

@ARTICLE{schultz2022,
       author = {{Schultz}, William C. and {Bildsten}, Lars and {Jiang}, Yan-Fei},
        title = "{Stochastic Low-frequency Variability in Three-dimensional Radiation Hydrodynamical Models of Massive Star Envelopes}",
      journal = {\apjl},
         year = 2022,
        month = jan,
       volume = {924},
       number = {1},
          eid = {L11},
        pages = {L11},
          doi = {10.3847/2041-8213/ac441f},
archivePrefix = {arXiv},
       eprint = {2110.13944},
 primaryClass = {astro-ph.SR},
       adsurl = {https://ui.adsabs.harvard.edu/abs/2022ApJ...924L..11S}
}

@article{sundqvist2018,
  title={Atmospheric NLTE models for the spectroscopic analysis of blue stars with winds-IV. Porosity in physical and velocity space},
  author={Sundqvist, JO and Puls, J},
  journal={A\&A},
  volume={619},
  pages={A59},
  year={2018},
  publisher={EDP Sciences}
}

@inproceedings{langer1997,
  title={The Eddington limit in rotating massive stars},
  author={Langer, Norbert},
  booktitle={Luminous Blue Variables: Massive Stars in Transition},
  volume={120},
  pages={83},
  year={1997}
}

@ARTICLE{santolaya1997,
       author = {{Santolaya-Rey}, A.~E. and {Puls}, J. and {Herrero}, A.},
        title = "{Atmospheric NLTE-models for the spectroscopic analysis of luminous blue stars with winds.}",
      journal = {\aap},
         year = 1997,
        month = jul,
       volume = {323},
        pages = {488-512},
       adsurl = {https://ui.adsabs.harvard.edu/abs/1997A&A...323..488S}
}

@ARTICLE{puls2005,
       author = {{Puls}, J. and {Urbaneja}, M.~A. and {Venero}, R. and {Repolust}, T. and {Springmann}, U. and {Jokuthy}, A. and {Mokiem}, M.~R.},
        title = "{Atmospheric NLTE-models for the spectroscopic analysis of blue stars with winds. II. Line-blanketed models}",
      journal = {\aap},
         year = 2005,
        month = may,
       volume = {435},
       number = {2},
        pages = {669-698},
          doi = {10.1051/0004-6361:20042365},
archivePrefix = {arXiv},
       eprint = {astro-ph/0411398},
 primaryClass = {astro-ph},
       adsurl = {https://ui.adsabs.harvard.edu/abs/2005A&A...435..669P}
}

@ARTICLE{hillier1998,
       author = {{Hillier}, D. John and {Miller}, D.~L.},
        title = "{The Treatment of Non-LTE Line Blanketing in Spherically Expanding Outflows}",
      journal = {\apj},
         year = 1998,
        month = mar,
       volume = {496},
       number = {1},
        pages = {407-427},
          doi = {10.1086/305350},
       adsurl = {https://ui.adsabs.harvard.edu/abs/1998ApJ...496..407H}
}

@INPROCEEDINGS{hillier2001,
       author = {{Hillier}, D. John and {Lanz}, Thierry},
        title = "{CMFGEN: A non-LTE Line-Blanketed Radiative Transfer Code for Modeling Hot Stars with Stellar Winds}",
    booktitle = {Spectroscopic Challenges of Photoionized Plasmas},
         year = 2001,
       editor = {{Ferland}, Gary and {Savin}, Daniel Wolf},
       series = {Astronomical Society of the Pacific Conference Series},
       volume = {247},
        month = jan,
        pages = {343},
       adsurl = {https://ui.adsabs.harvard.edu/abs/2001ASPC..247..343H}
}

@ARTICLE{grafener2002,
       author = {{Gr{\"a}fener}, G. and {Koesterke}, L. and {Hamann}, W. -R.},
        title = "{Line-blanketed model atmospheres for WR stars}",
      journal = {\aap},
         year = 2002,
        month = may,
       volume = {387},
        pages = {244-257},
          doi = {10.1051/0004-6361:20020269},
       adsurl = {https://ui.adsabs.harvard.edu/abs/2002A&A...387..244G}
}

@ARTICLE{hamann2004,
       author = {{Hamann}, W. -R. and {Gr{\"a}fener}, G.},
        title = "{Grids of model spectra for WN stars, ready for use}",
      journal = {\aap},
         year = 2004,
        month = nov,
       volume = {427},
        pages = {697-704},
          doi = {10.1051/0004-6361:20040506},
       adsurl = {https://ui.adsabs.harvard.edu/abs/2004A&A...427..697H}
}

@ARTICLE{sander2012,
       author = {{Sander}, A. and {Hamann}, W. -R. and {Todt}, H.},
        title = "{The Galactic WC stars. Stellar parameters from spectral analyses indicate a new evolutionary sequence}",
      journal = {\aap},
         year = 2012,
        month = apr,
       volume = {540},
          eid = {A144},
        pages = {A144},
          doi = {10.1051/0004-6361/201117830},
archivePrefix = {arXiv},
       eprint = {1201.6354},
 primaryClass = {astro-ph.SR},
       adsurl = {https://ui.adsabs.harvard.edu/abs/2012A&A...540A.144S}
}

@ARTICLE{howarth1997,
       author = {{Howarth}, Ian D. and {Siebert}, Kaj W. and {Hussain}, Gaitee A.~J. and {Prinja}, Raman K.},
        title = "{Cross-correlation characteristics of OB stars from IUE spectroscopy}",
      journal = {\mnras},
         year = 1997,
        month = jan,
       volume = {284},
       number = {2},
        pages = {265-285},
          doi = {10.1093/mnras/284.2.265},
       adsurl = {https://ui.adsabs.harvard.edu/abs/1997MNRAS.284..265H}
}

@ARTICLE{schmidth2009,
       author = {{Schmidt}, W. and {Federrath}, C. and {Hupp}, M. and {Kern}, S. and {Niemeyer}, J.~C.},
        title = "{Numerical simulations of compressively driven interstellar turbulence. I. Isothermal gas}",
      journal = {\aap},
         year = 2009,
        month = jan,
       volume = {494},
       number = {1},
        pages = {127-145},
          doi = {10.1051/0004-6361:200809967},
archivePrefix = {arXiv},
       eprint = {0809.1321},
 primaryClass = {astro-ph},
       adsurl = {https://ui.adsabs.harvard.edu/abs/2009A&A...494..127S}
}

@ARTICLE{hunter_2007,
       author = {{Hunter}, John D.},
        title = "{Matplotlib: A 2D Graphics Environment}",
      journal = {Computing in Science and Engineering},
         year = 2007,
        month = may,
       volume = {9},
       number = {3},
        pages = {90-95},
          doi = {10.1109/MCSE.2007.55},
       adsurl = {https://ui.adsabs.harvard.edu/abs/2007CSE.....9...90H}
}

@ARTICLE{harris_2020,
       author = {{Harris}, Charles R. and {Millman}, K. Jarrod and {van der Walt}, St{\'e}fan J. and {Gommers}, Ralf and {Virtanen}, Pauli and {Cournapeau}, David and {Wieser}, Eric and {Taylor}, Julian and {Berg}, Sebastian and {Smith}, Nathaniel J. and {Kern}, Robert and {Picus}, Matti and {Hoyer}, Stephan and {van Kerkwijk}, Marten H. and {Brett}, Matthew and {Haldane}, Allan and {del R{\'\i}o}, Jaime Fern{\'a}ndez and {Wiebe}, Mark and {Peterson}, Pearu and {G{\'e}rard-Marchant}, Pierre and {Sheppard}, Kevin and {Reddy}, Tyler and {Weckesser}, Warren and {Abbasi}, Hameer and {Gohlke}, Christoph and {Oliphant}, Travis E.},
        title = "{Array programming with NumPy}",
      journal = {\nat},
         year = 2020,
        month = sep,
       volume = {585},
       number = {7825},
        pages = {357-362},
          doi = {10.1038/s41586-020-2649-2},
archivePrefix = {arXiv},
       eprint = {2006.10256},
 primaryClass = {cs.MS},
       adsurl = {https://ui.adsabs.harvard.edu/abs/2020Natur.585..357H}
}

@ARTICLE{virtanen_2020,
       author = {{Virtanen}, Pauli and {Gommers}, Ralf and {Oliphant}, Travis E. and {Haberland}, Matt and {Reddy}, Tyler and {Cournapeau}, David and {Burovski}, Evgeni and {Peterson}, Pearu and {Weckesser}, Warren and {Bright}, Jonathan and {van der Walt}, St{\'e}fan J. and {Brett}, Matthew and {Wilson}, Joshua and {Millman}, K. Jarrod and {Mayorov}, Nikolay and {Nelson}, Andrew R.~J. and {Jones}, Eric and {Kern}, Robert and {Larson}, Eric and {Carey}, C.~J. and {Polat}, {\.I}lhan and {Feng}, Yu and {Moore}, Eric W. and {VanderPlas}, Jake and {Laxalde}, Denis and {Perktold}, Josef and {Cimrman}, Robert and {Henriksen}, Ian and {Quintero}, E.~A. and {Harris}, Charles R. and {Archibald}, Anne M. and {Ribeiro}, Ant{\^o}nio H. and {Pedregosa}, Fabian and {van Mulbregt}, Paul and {SciPy 1. 0 Contributors}},
        title = "{SciPy 1.0: fundamental algorithms for scientific computing in Python}",
      journal = {Nature Methods},
         year = 2020,
        month = feb,
       volume = {17},
        pages = {261-272},
          doi = {10.1038/s41592-019-0686-2},
archivePrefix = {arXiv},
       eprint = {1907.10121},
 primaryClass = {cs.MS},
       adsurl = {https://ui.adsabs.harvard.edu/abs/2020NatMe..17..261V}
}

@article{keppens_2020,
title = {MPI-AMRVAC: A parallel, grid-adaptive PDE toolkit},
journal = {Computers \& Mathematics with Applications},
volume = {81},
pages = {316-333},
year = {2021},
issn = {0898-1221},
doi = {https://doi.org/10.1016/j.camwa.2020.03.023},
url = {https://www.sciencedirect.com/science/article/pii/S0898122120301279},
author = {Rony Keppens and Jannis Teunissen and Chun Xia and Oliver Porth}
}

@ARTICLE{schultz2023,
       author = {{Schultz}, William C. and {Bildsten}, Lars and {Jiang}, Yan-Fei},
        title = "{Turbulence-supported Massive Star Envelopes}",
      journal = {\apjl},
         year = 2023,
        month = jul,
       volume = {951},
       number = {2},
          eid = {L42},
        pages = {L42},
          doi = {10.3847/2041-8213/acdf50},
archivePrefix = {arXiv},
       eprint = {2306.08034},
 primaryClass = {astro-ph.SR},
       adsurl = {https://ui.adsabs.harvard.edu/abs/2023ApJ...951L..42S}
}

@ARTICLE{gonzaleztora2025a,
       author = {{Gonz{\'a}lez-Tor{\`a}}, G. and {Sander}, A.~A.~C. and {Sundqvist}, J.~O. and {Debnath}, D. and {Delbroek}, L. and {Josiek}, J. and {Lefever}, R.~R. and {Moens}, N. and {Van der Sijpt}, C. and {Verhamme}, O.},
        title = "{Improving 1D stellar atmosphere models with insights from multi-dimensional simulations: I. 1D versus 2D stratifications and spectral comparisons for O stars}",
      journal = {\aap},
         year = {2025a},
        month = feb,
       volume = {694},
          eid = {A269},
        pages = {A269},
          doi = {10.1051/0004-6361/202452241},
archivePrefix = {arXiv},
       eprint = {2501.14511},
 primaryClass = {astro-ph.SR},
       adsurl = {https://ui.adsabs.harvard.edu/abs/2025A&A...694A.269G}
}

@ARTICLE{vandersijpt2025,
       author = {{Van der Sijpt}, C. and {Sundqvist}, J.~O. and {Debnath}, D. and {Driessen}, F.~A. and {Moens}, N.},
        title = "{Structure formation in O-type stars and Wolf─Rayet stars}",
      journal = {\aap},
         year = 2025,
        month = feb,
       volume = {694},
          eid = {A54},
        pages = {A54},
          doi = {10.1051/0004-6361/202452385},
archivePrefix = {arXiv},
       eprint = {2501.09378},
 primaryClass = {astro-ph.SR},
       adsurl = {https://ui.adsabs.harvard.edu/abs/2025A&A...694A..54V}
}

@ARTICLE{moens2025,
       author = {{Moens}, N. and {Debnath}, D. and {Verhamme}, O. and {Backs}, F. and {Van der Sijpt}, C. and {Sundqvist}, J.~O. and {Sander}, A.~A.~C.},
        title = "{2D unified atmosphere and wind simulations for a grid of O-type stars}",
      journal = {\aap},
         year = 2025,
        month = dec,
       volume = {704},
          eid = {A121},
        pages = {A121},
          doi = {10.1051/0004-6361/202556825},
archivePrefix = {arXiv},
       eprint = {2511.04542},
 primaryClass = {astro-ph.SR},
       adsurl = {https://ui.adsabs.harvard.edu/abs/2025A&A...704A.121M}
}

@ARTICLE{kida1990,
       author = {{Kida}, Shigeo and {Orszag}, Steven A.},
        title = "{Energy and spectral dynamics in forced compressible turbulence}",
      journal = {Journal of Scientific Computing},
         year = 1990,
        month = jun,
       volume = {5},
        pages = {85-125},
       adsurl = {https://ui.adsabs.harvard.edu/abs/1990JSCom...5...85K}
}

@ARTICLE{schillemans2026,
       author = {{Schillemans}, P. and {Sundqvist}, J.~O. and {Debnath}, D. and {Delbroek}, L. and {Moens}, N. and {Van der Sijpt}, C.},
        title = "{Simulations of massive star atmospheres and winds during giant eruptive and quiescent luminous blue variable phases}",
      journal = {arXiv e-prints},
         year = 2026,
        month = mar,
          eid = {arXiv:2603.22168},
        pages = {arXiv:2603.22168},
          doi = {10.48550/arXiv.2603.22168},
archivePrefix = {arXiv},
       eprint = {2603.22168},
 primaryClass = {astro-ph.SR},
       adsurl = {https://ui.adsabs.harvard.edu/abs/2026arXiv260322168S}
}

@ARTICLE{delbroek2025,
       author = {{Delbroek}, L. and {Sundqvist}, J.~O. and {Debnath}, D. and {Moens}, N. and {Backs}, F. and {Van der Sijpt}, C. and {Verhamme}, O. and {Schillemans}, P.},
        title = "{Spectral synthesis of 3D unified model atmospheres with winds for O stars}",
      journal = {\aap},
         year = 2025,
        month = dec,
       volume = {704},
          eid = {A104},
        pages = {A104},
          doi = {10.1051/0004-6361/202556421},
archivePrefix = {arXiv},
       eprint = {2507.12057},
 primaryClass = {astro-ph.SR},
       adsurl = {https://ui.adsabs.harvard.edu/abs/2025A&A...704A.104D}
}

@ARTICLE{kraichnan1980,
       author = {{Kraichnan}, R.~H. and {Montgomery}, D.},
        title = "{REVIEW ARTICLE: Two-dimensional turbulence}",
      journal = {Reports on Progress in Physics},
         year = 1980,
        month = may,
       volume = {43},
       number = {5},
        pages = {547-619},
          doi = {10.1088/0034-4885/43/5/001},
       adsurl = {https://ui.adsabs.harvard.edu/abs/1980RPPh...43..547K}
}

@ARTICLE{kraichnan1967,
       author = {{Kraichnan}, Robert H.},
        title = "{Inertial Ranges in Two-Dimensional Turbulence}",
      journal = {Physics of Fluids},
         year = 1967,
        month = jul,
       volume = {10},
       number = {7},
        pages = {1417-1423},
          doi = {10.1063/1.1762301},
       adsurl = {https://ui.adsabs.harvard.edu/abs/1967PhFl...10.1417K}
}

@ARTICLE{fouxon2023,
       author = {{Fouxon}, Itzhak and {Kritsuk}, Alexei G. and {Mond}, Michael},
        title = "{Compressible two-dimensional turbulence: cascade reversal and sensitivity to imposed magnetic field}",
      journal = {New Journal of Physics},
         year = 2023,
        month = nov,
       volume = {25},
       number = {11},
          eid = {113005},
        pages = {113005},
          doi = {10.1088/1367-2630/ad0172},
archivePrefix = {arXiv},
       eprint = {2206.10367},
 primaryClass = {physics.flu-dyn},
       adsurl = {https://ui.adsabs.harvard.edu/abs/2023NJPh...25k3005F}
}

@ARTICLE{jiang2018,
       author = {{Jiang}, Yan-Fei and {Cantiello}, Matteo and {Bildsten}, Lars and {Quataert}, Eliot and {Blaes}, Omer and {Stone}, James},
        title = "{Outbursts of luminous blue variable stars from variations in the helium opacity}",
      journal = {\nat},
         year = 2018,
        month = sep,
       volume = {561},
       number = {7724},
        pages = {498-501},
          doi = {10.1038/s41586-018-0525-0},
archivePrefix = {arXiv},
       eprint = {1809.10187},
 primaryClass = {astro-ph.SR},
       adsurl = {https://ui.adsabs.harvard.edu/abs/2018Natur.561..498J}
}

@ARTICLE{joss1973,
       author = {{Joss}, P.~C. and {Salpeter}, E.~E. and {Ostriker}, J.~P.},
        title = "{On the ``Critical Luminosity'' in Stellar Interiors and Stellar Surface Boundary Conditions}",
      journal = {\apj},
         year = 1973,
        month = apr,
       volume = {181},
        pages = {429-438},
          doi = {10.1086/152060},
       adsurl = {https://ui.adsabs.harvard.edu/abs/1973ApJ...181..429J}
}

@ARTICLE{gonzaleztora2025b,
       author = {{Gonz{\'a}lez-Tor{\`a}}, G. and {Sander}, A.~A.~C. and {Moens}, N. and {Sundqvist}, J.~O. and {Debnath}, D. and {Delbroek}, L. and {Josiek}, J. and {Lefever}, R.~R. and {Van der Sijpt}, C. and {Verhamme}, O. and {Bernini-Peron}, M.},
        title = "{Improving 1D stellar atmosphere models with insights from multidimensional simulations: II. 1D versus 3D model comparison for Wolf-Rayet stars}",
      journal = {\aap},
         year = {2025b},
        month = jul,
       volume = {699},
          eid = {A244},
        pages = {A244},
          doi = {10.1051/0004-6361/202554125},
archivePrefix = {arXiv},
       eprint = {2505.16458},
 primaryClass = {astro-ph.SR},
       adsurl = {https://ui.adsabs.harvard.edu/abs/2025A&A...699A.244G}
}

@ARTICLE{grassitelli2015,
       author = {{Grassitelli}, L. and {Fossati}, L. and {Sim{\'o}n-Di{\'a}z}, S. and {Langer}, N. and {Castro}, N. and {Sanyal}, D.},
        title = "{Observational Consequences of Turbulent Pressure in the Envelopes of Massive Stars}",
      journal = {\apjl},
         year = 2015,
        month = jul,
       volume = {808},
       number = {1},
          eid = {L31},
        pages = {L31},
          doi = {10.1088/2041-8205/808/1/L31},
archivePrefix = {arXiv},
       eprint = {1507.03988},
 primaryClass = {astro-ph.SR},
       adsurl = {https://ui.adsabs.harvard.edu/abs/2015ApJ...808L..31G}
}

@ARTICLE{kolmogorov1941,
       author = {{Kolmogorov}, A.},
        title = "{The Local Structure of Turbulence in Incompressible Viscous Fluid for Very Large Reynolds' Numbers}",
      journal = {Akademiia Nauk SSSR Doklady},
         year = 1941,
        month = jan,
       volume = {30},
        pages = {301-305},
       adsurl = {https://ui.adsabs.harvard.edu/abs/1941DoSSR..30..301K}
}

@incollection{burgers1948,
  author    = {{Burgers}, J. M.},
  title     = "{A Mathematical Model Illustrating the Theory of Turbulence}",
  editor     = {{von Mises}, R. and {von K{\'a}rm{\'a}n}, T.},
  booktitle  = "{Advances in Applied Mechanics}",
  volume      = {1},
  pages       = {171--199},
  year        = {1948},
  publisher   = {Academic Press},
  address     = {New York},
  doi         = {10.1016/S0065-2156(08)70100-5},
  adsurl      = {https://ui.adsabs.harvard.edu/abs/1948AdApM...1..171B}
}

@ARTICLE{kritsuk2007,
       author = {{Kritsuk}, Alexei G. and {Norman}, Michael L. and {Padoan}, Paolo and {Wagner}, Rick},
        title = "{The Statistics of Supersonic Isothermal Turbulence}",
      journal = {\apj},
         year = 2007,
        month = aug,
       volume = {665},
       number = {1},
        pages = {416-431},
          doi = {10.1086/519443},
archivePrefix = {arXiv},
       eprint = {0704.3851},
 primaryClass = {astro-ph},
       adsurl = {https://ui.adsabs.harvard.edu/abs/2007ApJ...665..416K}
}

@ARTICLE{goldberg2026,
       author = {{Goldberg}, Jared A. and {Jiang}, Yan-Fei and {Bildsten}, Lars and {Cantiello}, Matteo},
        title = "{Type IIb Supernova Progenitors in 3D: Variability and Episodic Mass Loss Revealed by Radiation Hydrodynamics Simulations}",
      journal = {\apjl},
         year = 2026,
        month = feb,
       volume = {998},
       number = {1},
          eid = {L10},
        pages = {L10},
          doi = {10.3847/2041-8213/ae37ec},
archivePrefix = {arXiv},
       eprint = {2508.12486},
 primaryClass = {astro-ph.SR},
       adsurl = {https://ui.adsabs.harvard.edu/abs/2026ApJ...998L..10G}
}

@ARTICLE{federrath2021,
       author = {{Federrath}, Christoph and {Klessen}, Ralf S. and {Iapichino}, Luigi and {Beattie}, James R.},
        title = "{The sonic scale of interstellar turbulence}",
      journal = {Nature Astronomy},
         year = 2021,
        month = jan,
       volume = {5},
        pages = {365-371},
          doi = {10.1038/s41550-020-01282-z},
archivePrefix = {arXiv},
       eprint = {2011.06238},
 primaryClass = {astro-ph.GA},
       adsurl = {https://ui.adsabs.harvard.edu/abs/2021NatAs...5..365F}
}

@ARTICLE{blaes2011,
       author = {{Blaes}, Omer and {Krolik}, Julian H. and {Hirose}, Shigenobu and {Shabaltas}, Natalia},
        title = "{Dissipation and Vertical Energy Transport in Radiation-dominated Accretion Disks}",
      journal = {\apj},
         year = 2011,
        month = jun,
       volume = {733},
       number = {2},
          eid = {110},
        pages = {110},
          doi = {10.1088/0004-637X/733/2/110},
archivePrefix = {arXiv},
       eprint = {1103.5052},
 primaryClass = {astro-ph.HE},
       adsurl = {https://ui.adsabs.harvard.edu/abs/2011ApJ...733..110B}
}

@book{pope2000,
  author    = {Pope, Stephen B.},
  title     = {Turbulent Flows},
  year      = {2000},
  publisher = {Cambridge University Press},
  address   = {Cambridge},
  doi       = {10.1017/CBO9780511840531}
}

\appendix 

\section{Numerical viscosity}
\label{app:visc}
\begin{figure}
     \centering
     \includegraphics[width=\linewidth]{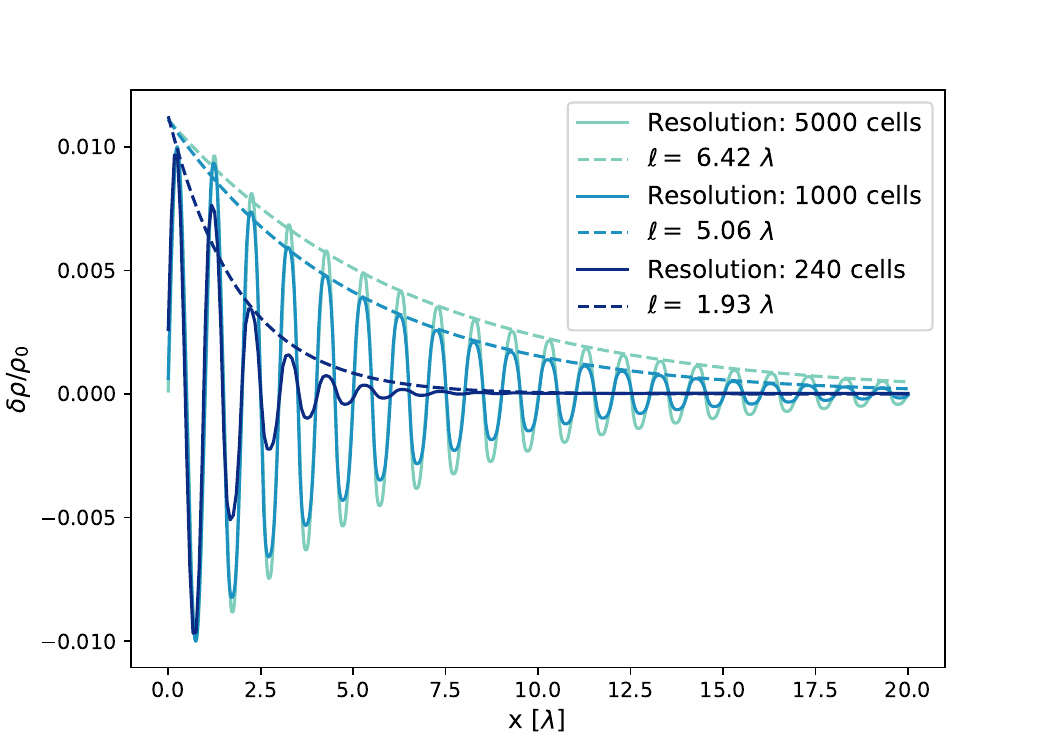}
     \caption{1D RHD wave at different resolutions.}
     \label{fig:wave_damping}
\end{figure}

\begin{figure}
     \centering
     \includegraphics[width=\linewidth]{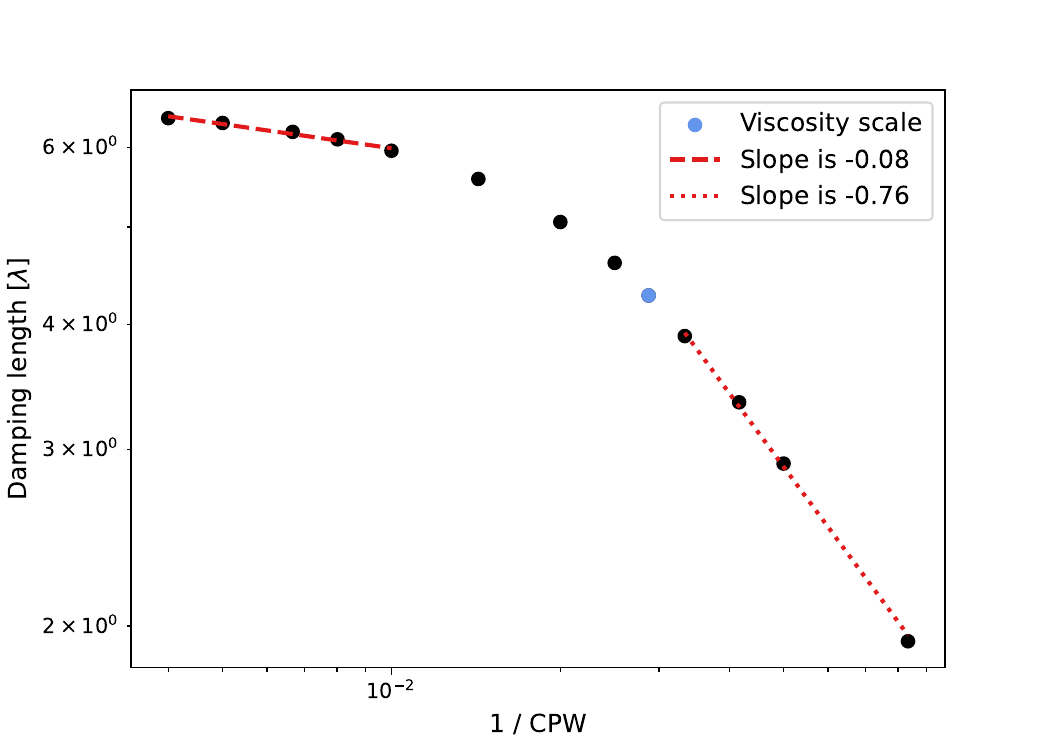}
     \caption{RHD wave damping lengths as a function of wavenumber.}
     \label{fig:damping_length_powerlaw}
\end{figure}

Since the radiation-hydrodynamics (RHD) equations solved in this work do not have a viscosity term, the dissipation scale is set entirely by the numerical viscosity. To estimate the spatial scale at which the numerical viscosity has a significant effect, we use the 1D RHD wave test case in \citet{moens2021}, changing the numerical schemes to match those used for the 2D simulations. This test case simulates a forced acoustic wave in  a radiation-dominated environment. This wave will be damped due to rapid radiative diffusion and an analytical solution for the damping length can be found from a dispersion relation found by \citet{mihalas} (their Sec.~8.1). For the background values chosen here, this damping length should be $8.16 \lambda$, with $\lambda$ the wavelength of the induced wave. The original set-up in \citet{moens2021} uses 2000 points to resolve a domain of $20 \lambda$, although it is important to note that a different, less diffusive flux limiting scheme was used in the original test. By changing the resolution, we can estimate the scale at which numerical viscosity becomes significant for our specific combination of numerical schemes. 
In Fig.~\ref{fig:wave_damping}, we show three wave solutions with different resolutions. The damping length is then derived by fitting an exponential function to the wave peaks. In this way, we can find the damping lengths as a function of the number of cells per wavelength (CPW). Since the domain and wavelength are always kept constant at $20 \lambda$, for each varying number of cells $N$, we get $\mathrm{CPW} = N / 20$. The CPW can easily be translated to our full RHD setup where we now keep the number of cells constant and vary the wavelength of the structures as we go to smaller scales. Generally, we can calculate $\mathrm{CPW} = \lambda / \Delta x$, where $\Delta x$ is the cell size, which is constant. Therefore, we know that $\mathrm{CPW} \sim \lambda$ and thus, for wavenumber $k$, $k \sim 1 / \mathrm{CPW}$. Fig.~\ref{fig:damping_length_powerlaw} shows the damping lengths as a function of wavenumber. Firstly, we note that all wavenumbers experience numerical dissipation to some degree. However, there is a clear cut-off wavenumber after which the decay in damping lengths becomes significantly stronger. We find that, for wavenumbers smaller than this cut-off, $\ell \sim k^{-0.08}$ and, for the larger wavenumbers, $\ell \sim k^{-0.76}$. Because of this change in slope, we take the cut-off $\mathrm{CPW} \approx 35$ as the numerical viscosity scale, indicated by a blue dot in the figure. This can be translated to a length scale in our 2D simulation setup as $\lambda = \mathrm{CPW} \Delta x$ with $\Delta x = 0.4 R_\mathrm{0} / 512$. Then we get $\lambda = 0.027R_\mathrm{0}$ which corresponds to a dissipation scale wavenumber $k_\mathrm{visc} \approx 230 / R_\mathrm{0}$.

\section{Resolution}
\label{app:resolution}
\begin{figure}
     \centering
     \includegraphics[width=\linewidth]{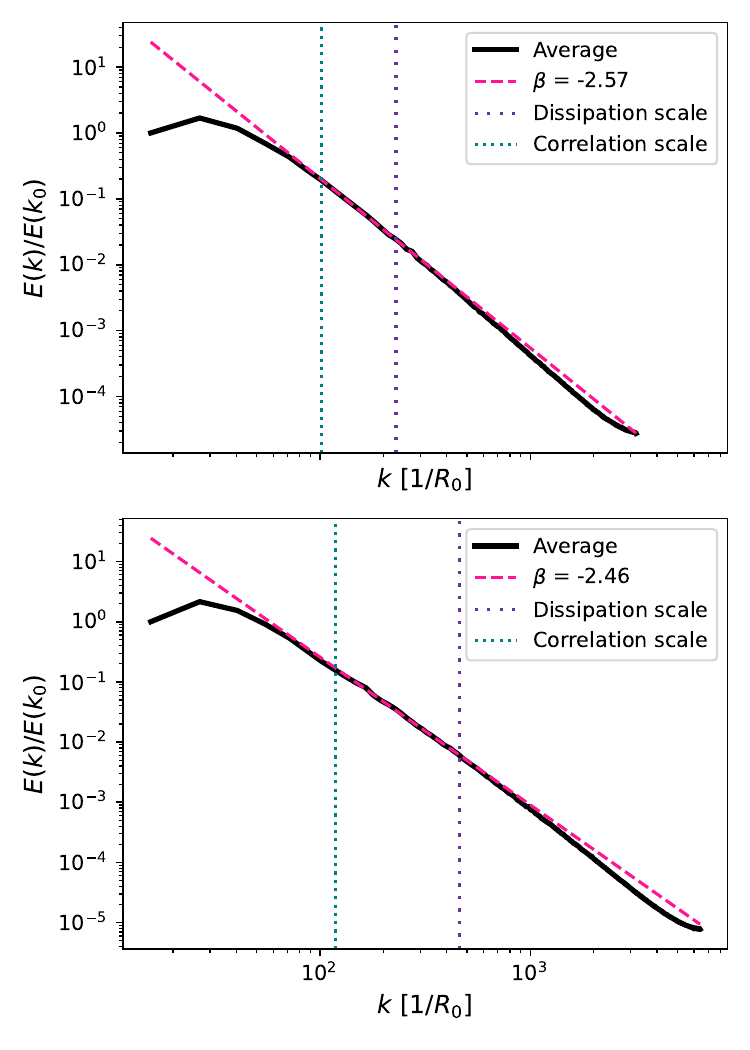}
     \caption{Comparison of energy power spectra for model G9 at different resolutions. The top panel shows the original resolution and the bottom panel shows the higher resolution. The green and purple dotted vertical lines indicate the correlation scale and the dissipation scale, respectively. The dashed lines are the power-law fits to the inertial range.}
     \label{fig:resolution_comparison}
 \end{figure}
The resolution of the grid determines the extent of the inertial range and can therefore impact the measured scaling exponent. Thus, it is important to check the convergence of this scaling exponent with increasing resolution. The original resolution of the simulations by \citet{moens2025}, at the highest refinement level, results in a cell size of $\Delta x = 8 \times 10^{-4} R_0$. As discussed in App.~\ref{app:visc}, this leads to a viscosity scale of $k_\mathrm{visc} \approx 230 / R_0$. As a convergence test, we also ran model G9 at higher resolution, doubling the number of cells and thus halving the cell size to $\Delta x = 4 \times 10^{-4} R_0$. This increases the viscosity cut-off wavenumber to $k_\mathrm{visc} \approx 460 / R_0$. In Fig.~\ref{fig:resolution_comparison}, we show the two averaged power spectra for G9 using a different grid resolution. The top panel shows the result from the original grid resolution and the lower panel shows the result after doubling the resolution. The green and purple dotted vertical lines show the correlation scale and the dissipation scale, respectively. The dashed lines are the power-law fits to the inertial range in each model. We see that the higher resolution model gives a scaling exponent of $\beta = -2.46$ and the lower resolution model gives $\beta = -2.57$. Since this is only a very minor difference between the two exponents, we conclude that the model resolution in \citet{moens2025} does not greatly impact the measured scaling exponents. 

\end{document}